\documentclass[]{spie}  
\usepackage[]{graphicx}
\usepackage[]{hyperref}

\title{Data reduction for the MATISSE instrument} 

\author{F. Millour\supit{a,}\footnote{\ \ O. Chesneau also contributed to the design and specifications of the MATISSE DRS, but passed away before seeing the result.},
           P. Berio\supit{a},
           M. Heininger\supit{b},
           K.-H. Hofmann\supit{b},
           D. Schertl\supit{b},
           G. Weigelt\supit{b},
           F. Guitton\supit{a},
           W. Jaffe\supit{c},
           U.~Beckmann\supit{b},
           R. Petrov\supit{a},
           F. Allouche\supit{a},
           S.~Robbe-Dubois\supit{a},
           S. Lagarde\supit{a},
           A.~Soulain\supit{a},
           A. Meilland\supit{a},
           A. Matter\supit{a},
           P. Cruzal\`ebes\supit{a},
           and~B.~Lopez\supit{a}.
\skiplinehalf
\supit{a}Universit\'e C\^ote d'Azur, OCA, CNRS, Lagrange, France; \\
\supit{b}Max Planck Institute for Radio Astronomy, Bonn, Germany;\\
\supit{c}Huygens Laboratory, Leiden, The Netherlands.
}

\authorinfo{
\vspace{0.cm} \\\emph{Further author information:}\\ Send correspondence to F. Millour, Lagrange, OCA, Bd de l'Observatoire, CS 34229, 06304, Nice, France\\E-mail: \href{mailto:fmillour@oca.eu}{fmillour@oca.eu}, Telephone: +33 (0)4 92 00 30 68, or +33 (0)4 92 07 64 89}

\begin{document} 
\maketitle 

\begin{abstract}
We present in this paper the general formalism and data processing
steps used in the MATISSE data reduction software, as it has been
developed by the MATISSE consortium. The MATISSE instrument is the
mid-infrared new generation interferometric instrument of the Very
Large Telescope Interferometer (VLTI). It is a 2-in-1 instrument with
2 cryostats and 2 detectors: one $2k\times2k$ Rockwell Hawaii 2RG
detector for L\&M-bands, and one $1k\times1k$ Raytheon Aquarius
detector for N-band, both read at high framerates, up to 30 frames per
second. MATISSE is undergoing its first tests in laboratory today.
\end{abstract}



\section{INTRODUCTION}
\label{sec:intro}  

\begin{figure}[htbp]
\begin{center}
\begin{tabular}{cc}
\includegraphics[width=0.35\textwidth]{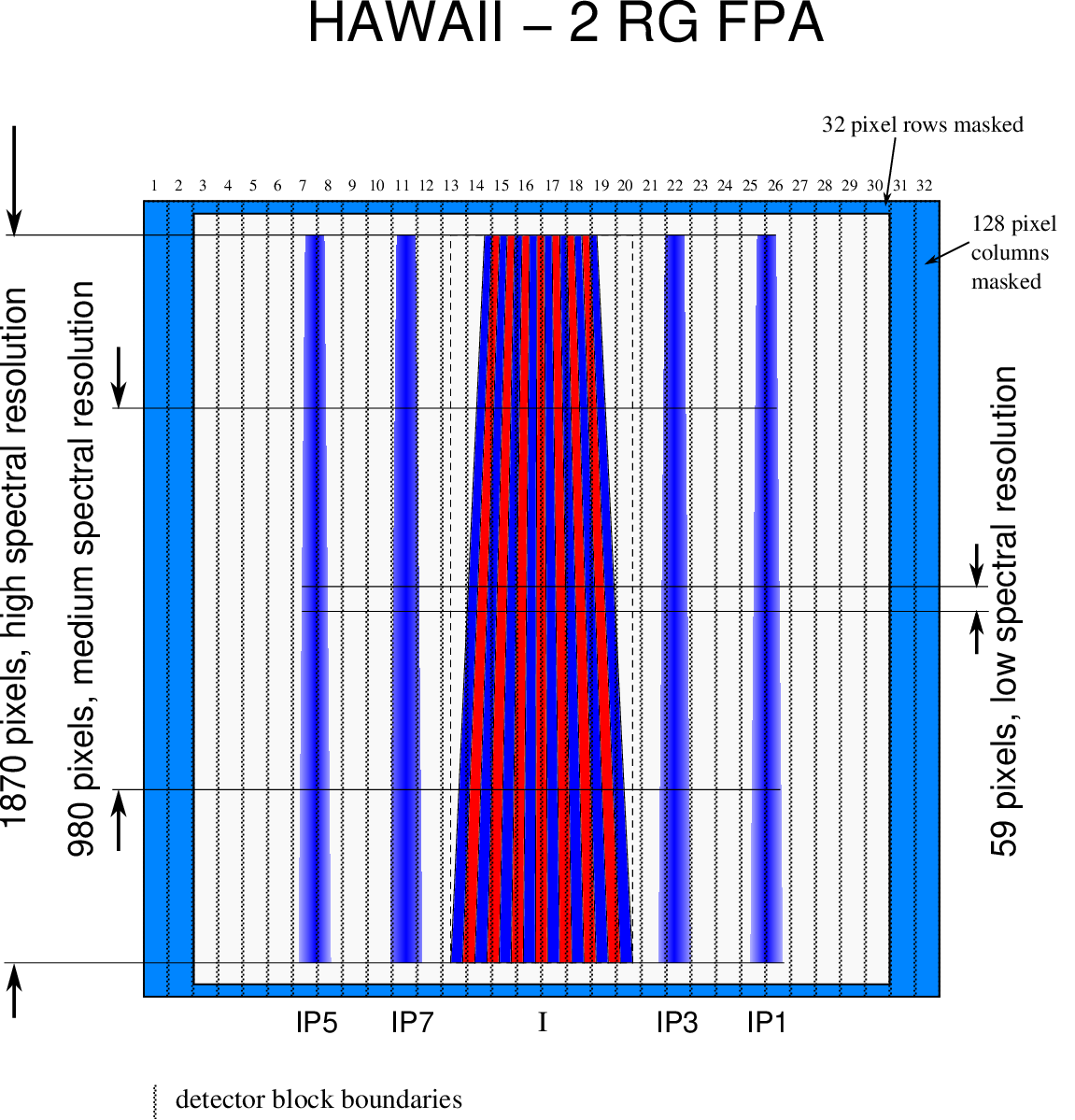}&
\includegraphics[width=0.52\textwidth]{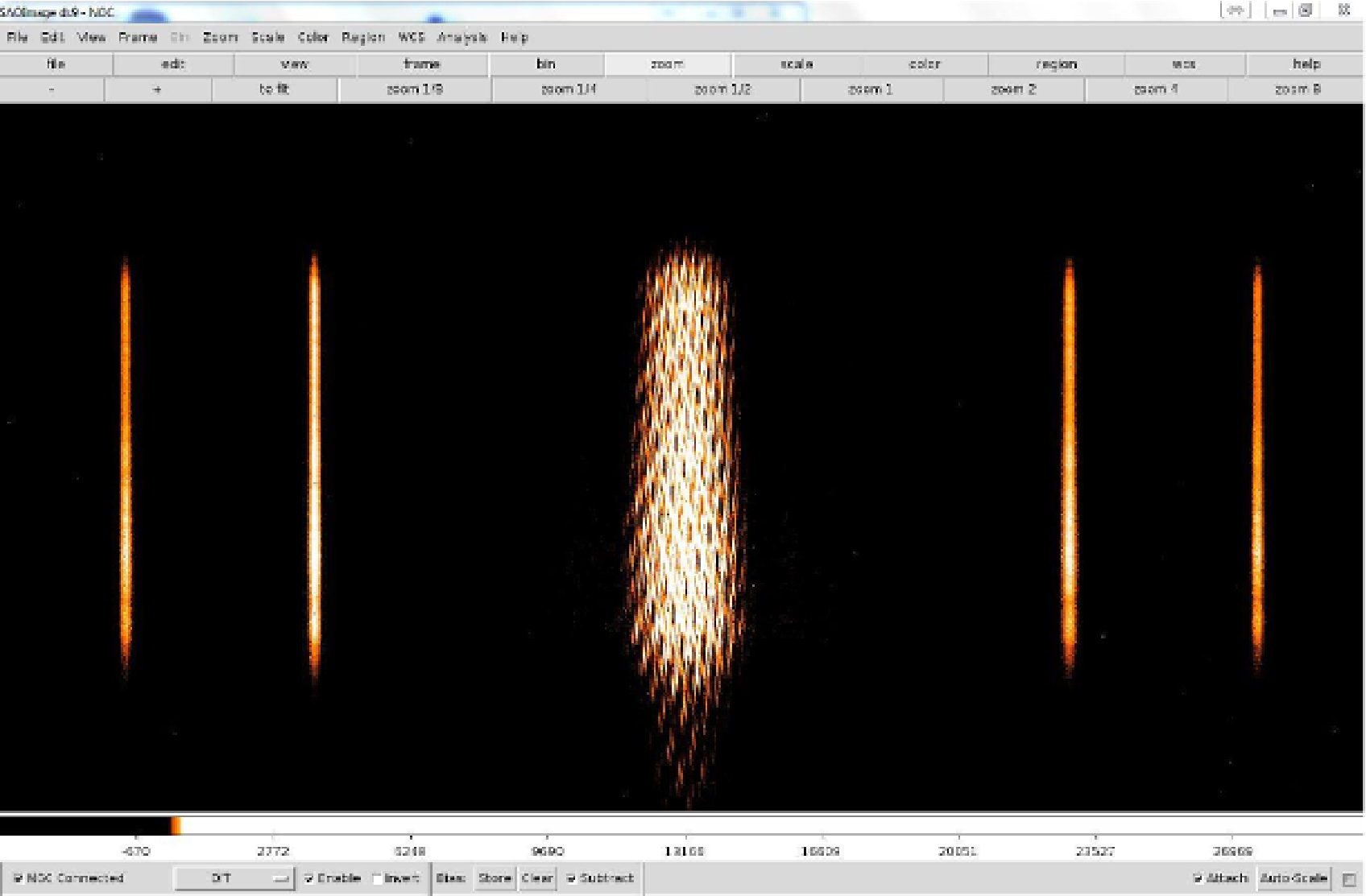}
\end{tabular}
\end{center}
\caption[Detector Fringes] 
{ \label{fig:fringes} Expected (Left) vs. real data (Right)
beams implementation on the MATISSE Hawaii detector
(L-band).  The wide central stripe contains the dispersed
fringes (vertical spectral dispersion), while the 4 narrow
stripes on the side contain the 4 single-telescope
photometry used for flux calibration.}
\end{figure}

MATISSE\cite{2014Msngr.157....5L} produces 4-telescopes interferences
which are dispersed onto an infrared detector with a spectrograph,
simultaneously in the L\&M-bands (3.5--4.1$\mu$m \& 4.6--5.5$\mu$m,
respectively), and in the N-band (8--13$\mu$m). Similarly as
AMBER\cite{2004SPIE.5491.1222M}, MATISSE uses an all-in-one multiaxial
combination with (SI-PHOT mode) or without (HIGH-SENS mode)
photometric separation (see Fig.~\ref{fig:fringes}), but due to the
limits of the P2VM algorithm that were found for the AMBER instrument,
we chose to use the classical Fourier processing instead of a P2VM
algorithm for the data reduction software (DRS).

The MATISSE interferogram $i(x, \lambda, t)$ is governed by the
interferometric equation, describing the signal:

\begin{eqnarray}
i(x,\lambda, t) & = & \sum_{i=1}^4 P_{i} + \sum_{i=1}^4\sum_{j=i+1}^4
2 V_{ij} \cdot \sqrt{P_{i} \cdot P_{j}} \cdot \cos\left( 2 \pi \cdot
f_{ij} \cdot x + \Phi_{ij}\right) + B
\label{eq:interferogram}
\end{eqnarray}

Most, if not all, terms of this equation depend on the coordinates $x$
(position on the detector in the space direction) and $\lambda$
(position on the detector in the spectral direction), and also time
$t$. However, for clarity, we exhibit their dependence only in each
terms description: $P_i(x,\lambda,t)$ is the flux received on the
detector from the telescope $i$. $V_{ij}(\lambda,t)$ denotes the
visibility of the observed object on the telescope pair $ij$ (constant
but wavelength-dependent) multiplied by an instrumental and
atmospheric transfer function (an unknown, but supposedly slowly
variable gain $\leq1$), and $\Phi_{ij}(\lambda,t)$ is composed of the
phase of the object (constant but wavelength-dependent) plus the
atmospheric phase errors (depending both on $\lambda$ and $t$). Let
$b_{ij}$ be the combining baseline, i.e. the separation of each of the
6 telescopes pair as seen by the detector, fixed by construction. The
fringe rate, or frequency, of the fringe pattern is directly linked to
it by $f_{ij}(\lambda) = \frac{b_{ij}}{\lambda}$. In the case of
MATISSE, we have $b_{ij}$ equal to $3\,D$, $6\,D$, $9\,D$, $12\,D$,
$15\,D$ and $18\,D$, $D$ being the dimension of the output pupil of
the instrument, right in front of the detector. Finally,
$B(x, \lambda, t)$ is a dominant and highly variable background
contribution from the warm and turbulent atmosphere, the warm optical
train of the VLTI and of MATISSE itself. $B(x, \lambda, t)$ can be up
to $10^5$ times larger than the flux from the star, and we will see
that mitigating this background contribution is a serious task in the
MATISSE concept and DRS (as it is for any mid-infrared instrument).

\section{THE DATA REDUCTION SOFTWARE} 

The MATISSE DRS (called \texttt{drsmat}) aims at producing
science-grade spectra, visibilities and phases out of the MATISSE raw
datasets. The description of the main algorithms were described in
details in several technical documents, and we report here the
selected data reduction strategy and algorithms that were implemented
in \texttt{drsmat}.

It is implemented in ANSI C and is fully integrated into the ESO
pipeline software (i.e. it is interfaced
with \texttt{EsoRex}\cite{2015ascl.soft04003E}
and \texttt{Reflex}\cite{2013A&A...559A..96F}). The implementation
uses the ESO-developped Common Pipeline Library
(CPL)\cite{2004SPIE.5493..444M}, the FITS format for intermediate
frames, and the OIFITS format\cite{2005PASP..117.1255P,
2015arXiv151004556D} for the final data products (the MATISSE DRS will
fully support the OIFITS version 2 format). The \texttt{Reflex}
interfaces are made with xml files and also Organisation,
Classification, and Association rules (OCA rules), and the GUI part is
coded in python. The consortium actively supports ESO in the
integration of \texttt{drsmat} in a \texttt{Reflex} workflow that will
be used both at ESO and at the consortium sides for the reduction of
the MATISSE data.


\subsection{Overview}

The basic input data are the raw frames of the MATISSE instrument,
stored as FITS files. Let be $i^{\rm raw}(X,Y, t)$ this raw data
intensity at time $t$, given at the $(X,Y)$ pixel detector
index. Compared to the interferogram $i(x, \lambda, t)$, the raw data
is affected by the following effects:
\begin{itemize}
\item Detector bias,
\item Detector bad pixels,
\item Detector + instrument flat field, plus eventually non-linear effects,
\item Instrument distortion map, including a non-linear spectral dispersion law,
\end{itemize}

\begin{figure}[htbp]
\begin{center}
\begin{tabular}{c}
\includegraphics[width=1\textwidth]{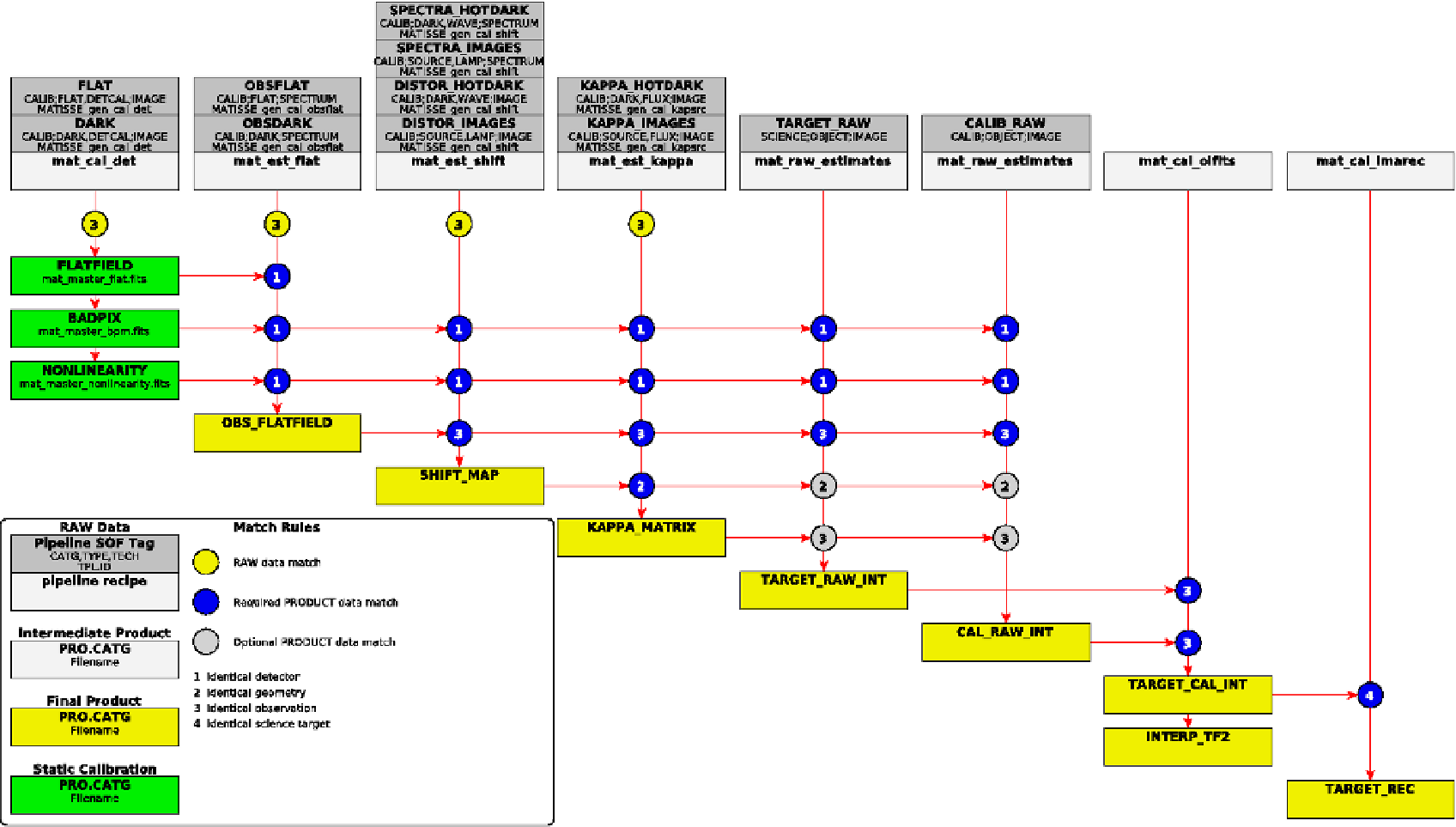}
\end{tabular}
\end{center}
\caption[Pipeline Cascade] 
{ \label{fig:pipeline_cascade} The MATISSE \texttt{drsmat}
pipeline cascade, showing all the steps and files necessary
to obtain a final science-grade dataset (calibrated oifits
and/or reconstructed images).  }
\end{figure}

All these effects must be removed from the data in order to get the
interferogram of eq.~\ref{eq:interferogram}, out of which we extract
the MATISSE observables. In addition, a calibration process leads to
science-grade data. To do so, the software is organized into a data
reduction cascade (shown in Fig.~\ref{fig:pipeline_cascade}) that
enables one to produce the calibration maps (\texttt{mat\_cal\_det},
\texttt{mat\_est\_flat}, \texttt{mat\_est\_shift},
\texttt{mat\_est\_kappa}), reduce the raw data into oifits files
(\texttt{mat\_raw\_estimates}), calibrate the data
(\texttt{mat\_cal\_oifits}), and then reconstruct an image out of the
reduced data (\texttt{mat\_cal\_imarec}).

The very first step is to estimate the calibration maps like flat
field map, bad pixels map, or distortion map out of calibration frames
obtained during daytime using the internal calibration source of the
instrument. They are used to correct the basic cosmetic of the
detector and some optical effects of the instrument.

\subsection{Detector-specific cosmetics} 

\subsubsection{Bad Pixels, Flat Field} 

The bad pixel map is derived from a series of cold dark raw frames
obtained by increasing the integration time (DIT) from the MINDIT to
five times the MINDIT with a minimum of 50 frames per exposure:

\begin{itemize}
\item First, spurious cosmics are detected by analyzing the intensity
values of each pixel,
\item Then, a robust median, a mean, and a standard deviation are computed
for each pixel.
\item For quality control purposes, the median, mean, and standard
deviation of each pixel median of each series are calculated.
\item For each pixel, a straight line is fitted to the 5 different
median values. This fit results in an offset (pixel bias), a slope
(dark current), and a fit quality.
\item A pixel is marked as a bad pixel if the fit quality or the
slope exceeds a certain limit (N sigmas of the standard deviation of
the fit quality or slope for all detector pixels).
\end{itemize}

The flatfield map is derived from a series of illuminated frames
ranging from detector noise up to detector saturation:

\begin{itemize}
\item First, the median of the matching series of cold dark frames is
subtracted.
\item Each frame is processed like the series of cold dark frames,
i.e., mean, median, and RMS are computed.
\item After normalization, the slopes of the straight lines from the
flatfield images are used to compute the master gain map. However, in
order to get a ``detector'' flatfield, the non-flat illumination must
be taken into account. Therefore, the implemented algorithm must be
adapted to the instrument optics.
\end{itemize}

The non-linearity map and the flat field are derived by fitting
higher-order polynomials to the exposed frames, and saved as maps into
fits files. We differentiate two types of flat fields in MATISSE:
the \emph{detector flat field}, which is obtained by a special device
put into the beam, which diffuses the light onto the detector in a
roughly uniform way, and the \emph{observing flat field}, which is
obtained through the spectrograph with a white lamp or a star without
spectral features. The first one enables us to correct for detector
pixel-to-pixel gain variations, while the second one allows us to
correct for optics transmission variations.

\subsubsection{Excess low frequency noise}

\begin{figure}[htbp]
\begin{center}
\begin{tabular}{cccc}
\includegraphics[width=0.6\textwidth]{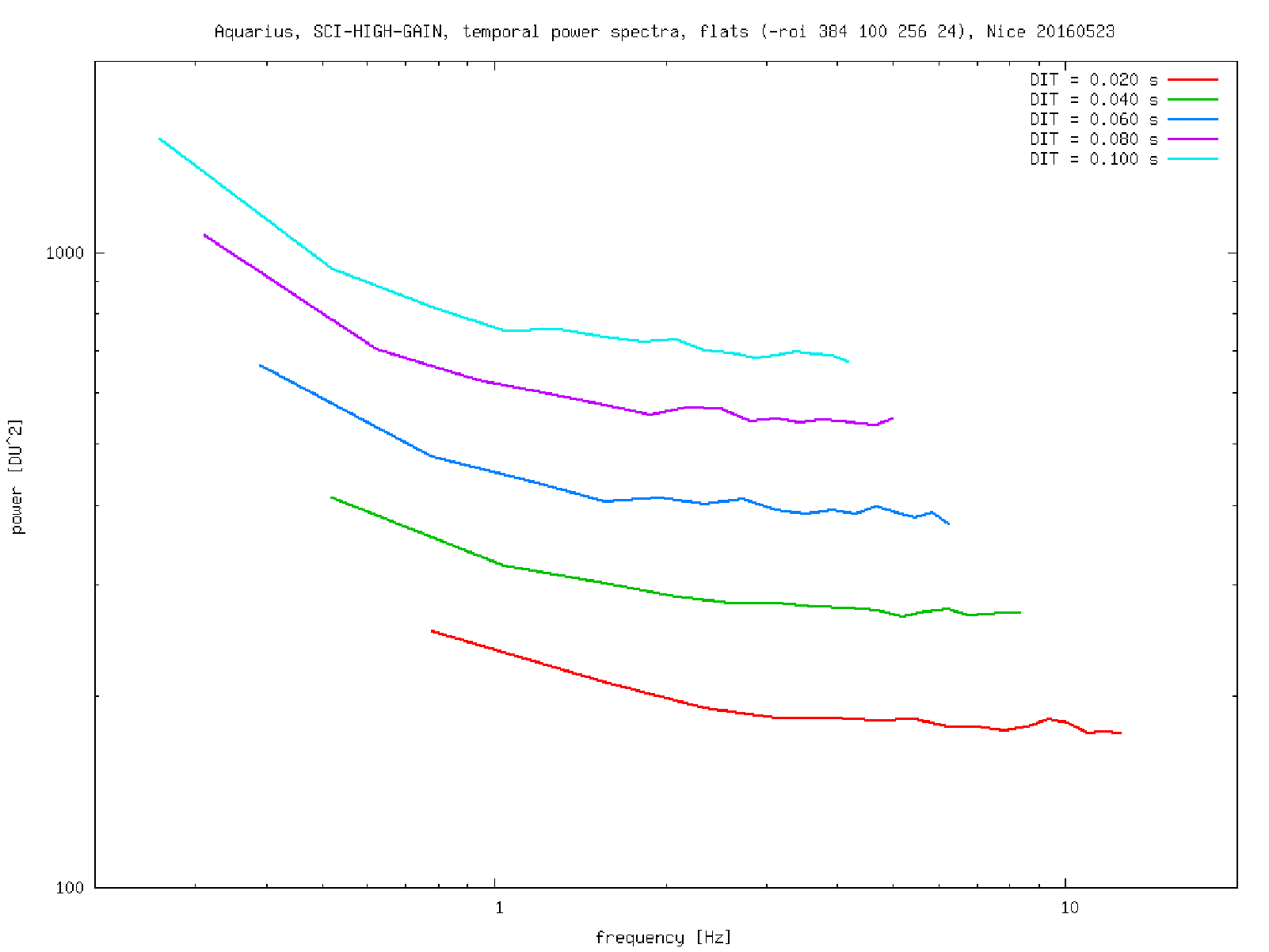}
\end{tabular}
\end{center}
\caption[Excess low frequency noise] 
{ \label{fig:ELFN} Average temporal power spectrum at
several exposure times and increasing illumination. The ELFN
shows up at frequencies lower than 1 Hz.}
\end{figure}

During the Aquarius detector tests, and thanks to the ESO experience
with VISIR, a strong temporal noise was effectively detected in the
data. A set of exposures with increasing DIT was taken. This so called
Excess Low Frequency Noise (ELFN) showed up in the temporal power
spectra, as presented in Fig.~\ref{fig:ELFN}. Pure white noise would
result in a nearly flat temporal power spectrum. The increase at lower
frequencies shows the ELFN.

Further investigations have lead to the following: the Aquarius ELFN
is not a simple 1/f noise, the origin of the ELFN seems to be inside
the semiconductor diode (pixel), the ELFN of different pixels is not
correlated, the ELFN could not be reduced by using different readout
modes, different multiplexer clocking schemes and delays or different
bias voltages, the ELFN is proportional to the detector output signal
level, ELFN cannot be seen in dark images, therefore CDS does not
help, and finally ELFN was found on other MIR detectors (Si:As, InSb,
etc.).

We expect this effect to be negligible for the coherent flux estimate
(eq.~\ref{eq:coherentFlux}), but likely present in the photometric
estimates of MATISSE (eq.~\ref{eq:KappaMatrix}). The mitigation of
this effect will need a fast chopping frequency like in VISIR, whose
speed need to be addressed on sky.

\subsection{Optics calibration} 

\subsubsection{Camera \& spectrograph image distortion}

Once the bad pixel map and flat field map have been produced, one can
continue the instrument calibration with the determination of the
image distortion. Since MATISSE contains a long-slit spectrograph, the
distortion in the wavelength direction translates into a spectral
dispersion law which is non-linear. We chose to treat the problem
globally by computing a ``shift map''.

\begin{figure}[htbp]
\begin{center}
\begin{tabular}{cccc}
\includegraphics[width=0.22\textwidth]{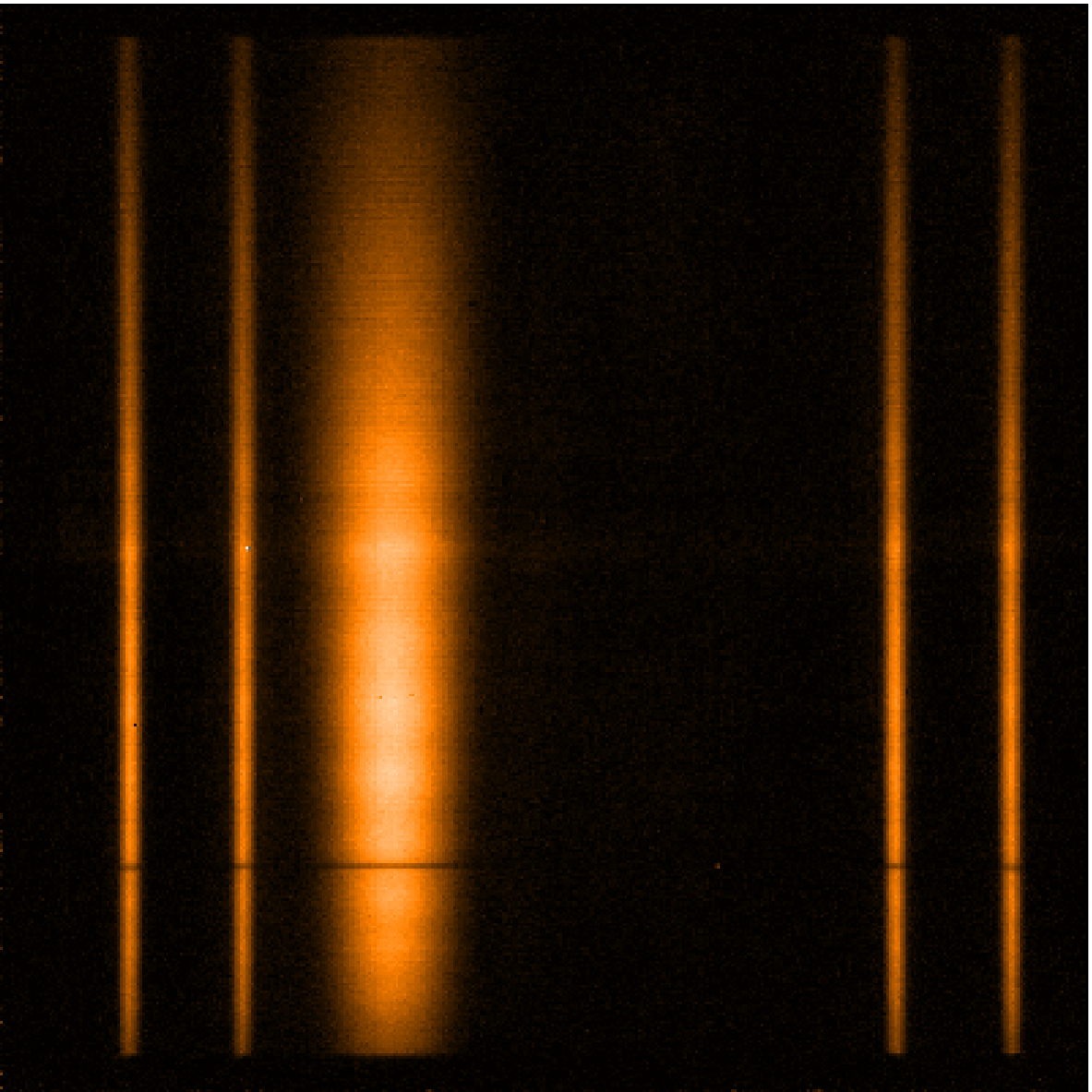}&
\includegraphics[width=0.22\textwidth]{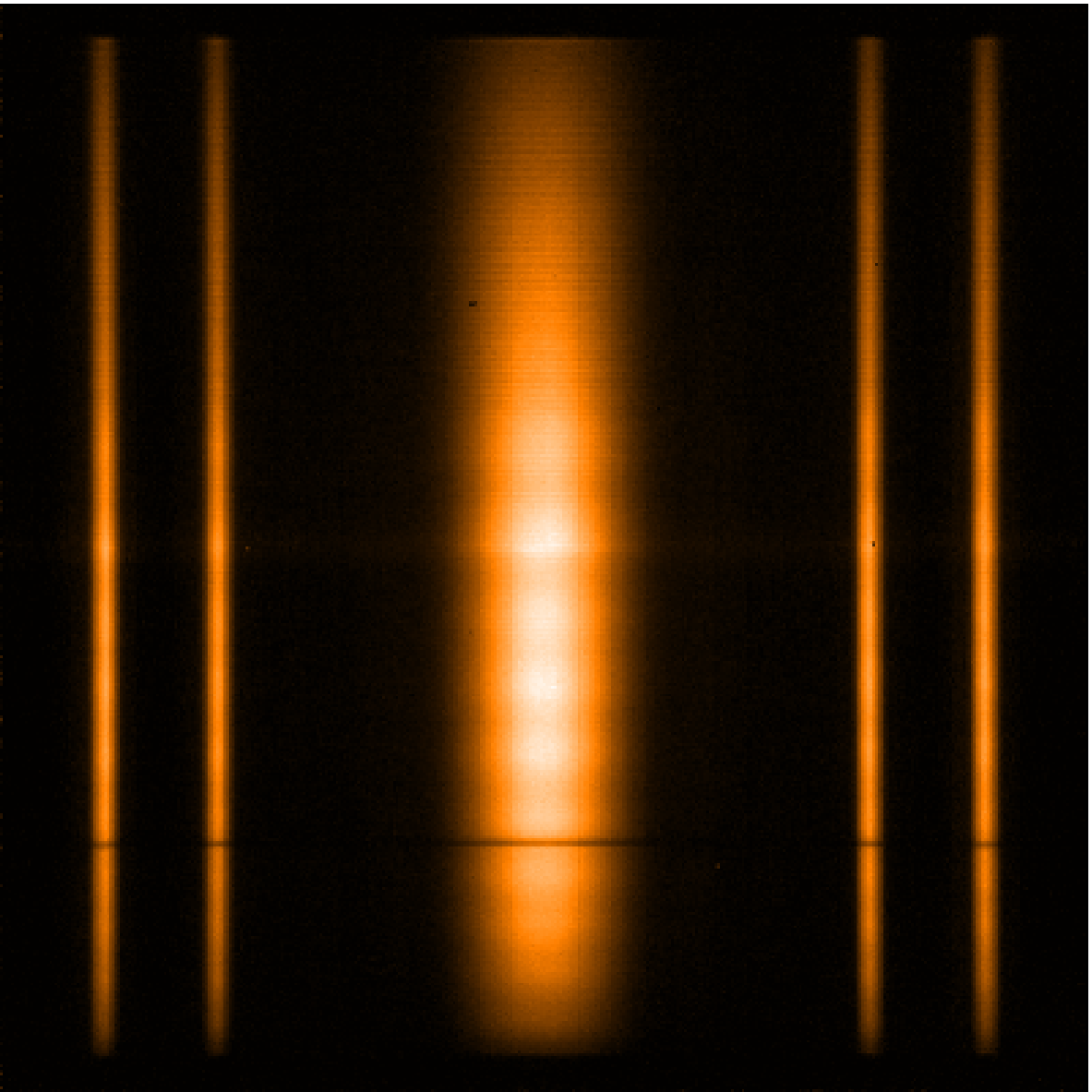}&
\includegraphics[width=0.22\textwidth]{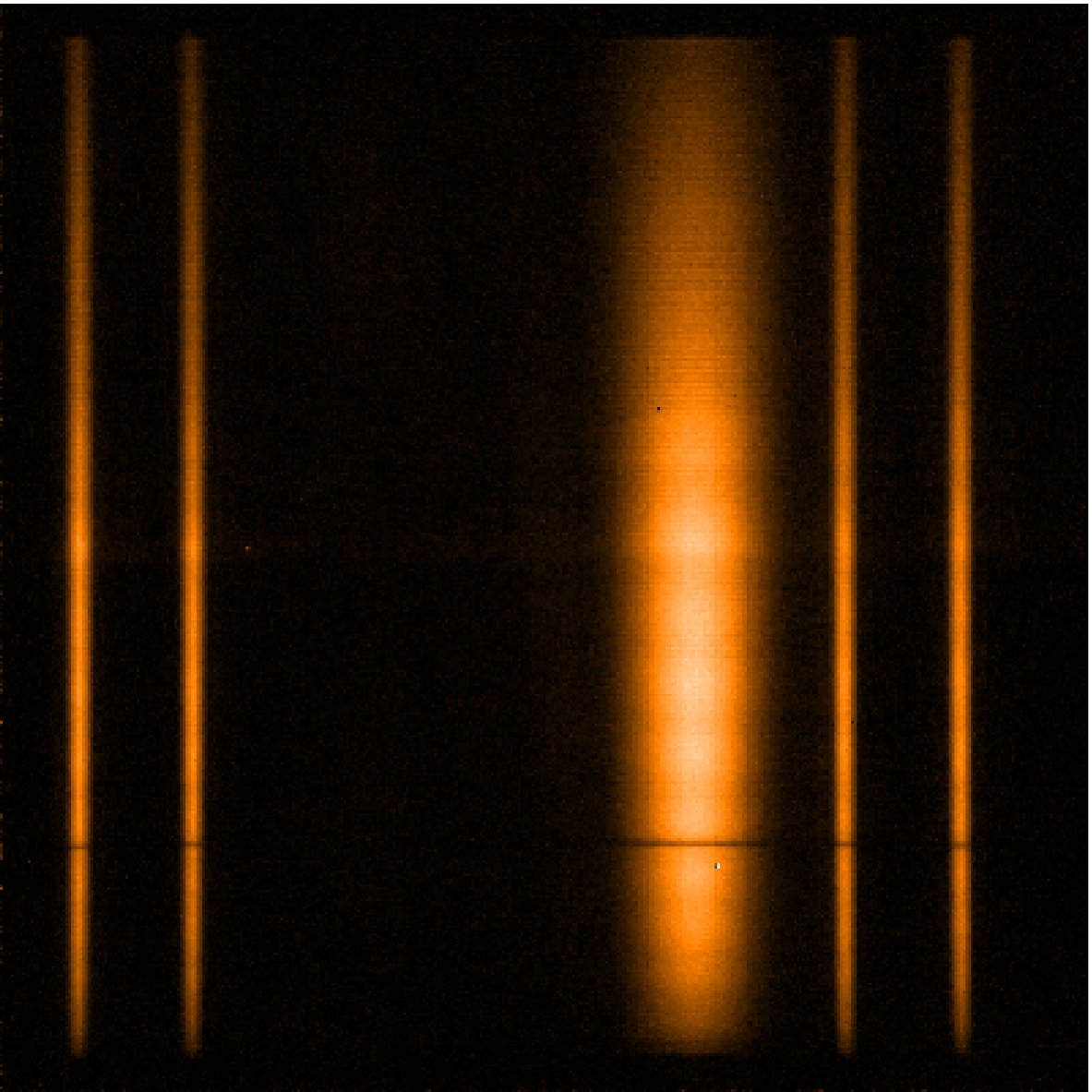}&
\includegraphics[width=0.22\textwidth]{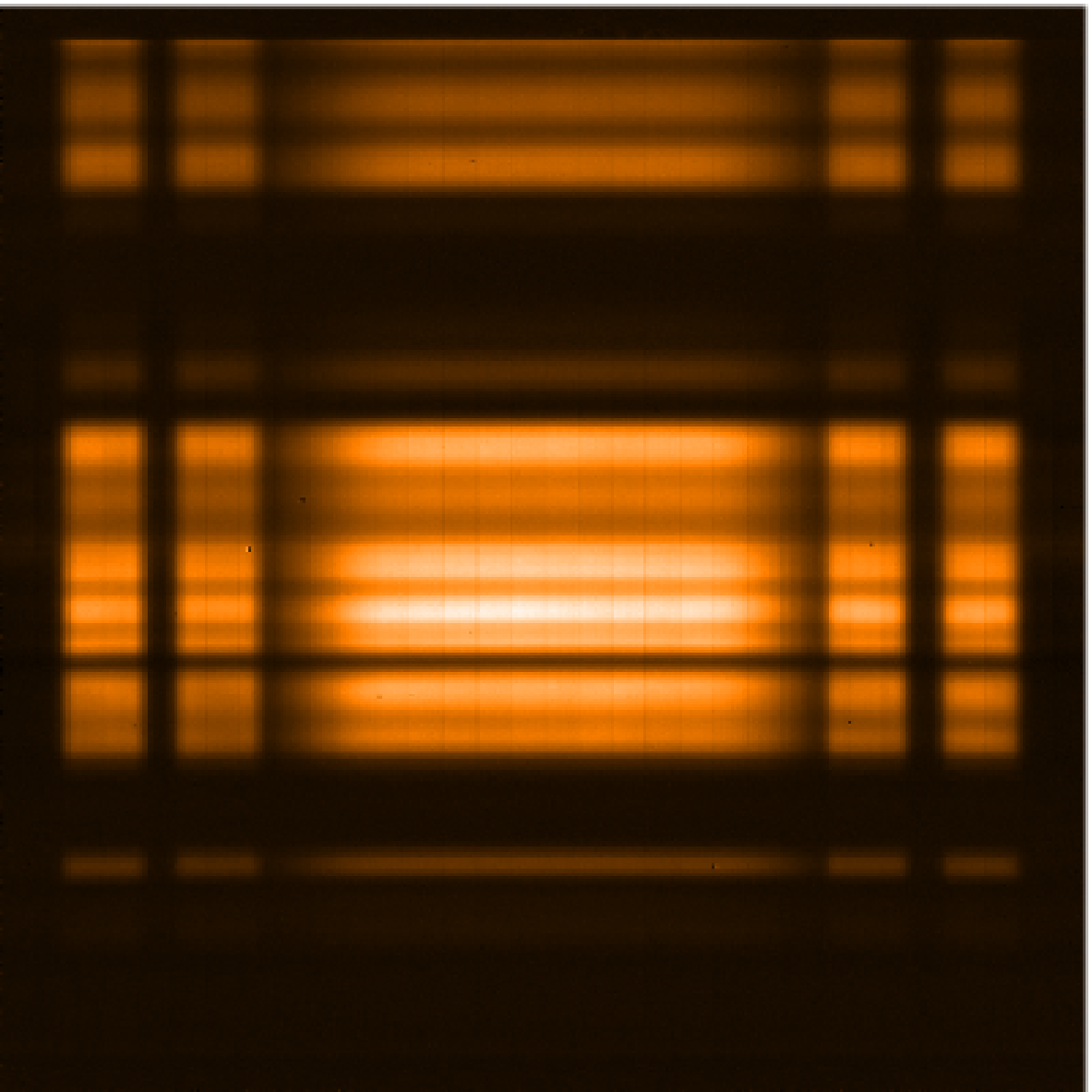}
\end{tabular}
\end{center}
\caption[Distortion Estimation] 
{ \label{fig:shift_map} {\bf Three panels on the left:}
Spatial grid, made with three pinholes at different
positions. Note that the photometric beams position is
inverted compared to the interferometric beam one. {\bf
Right:} Spectral grid, made with plastic foils with spectral
features. These snapshots were recorded in the N-band.}
\end{figure}

The shift map is derived from a series of frames containing either a
spatial grid (3 separate holes in the slit direction) or a spectral
grid introduced by carefully-chosen plastic foils (for the wavelength
direction), as shown in Fig.~\ref{fig:shift_map}. This distortion
estimate will be done at the same frequency as the flat field map or
bad pixel map estimates.

The distortion can be expressed as the relation between the pixel
coordinates (X, Y) and the spectrograph coordinates $(x, \lambda)$ on the
detector:
\begin{eqnarray}
X & = & a(x,\lambda) x + b(x,\lambda) \lambda \nonumber\\
Y & = & c(x,\lambda) x + d(x,\lambda) \lambda
\end{eqnarray}

$(X, Y)$ are the pixel indices, and $(x, \lambda)$ are the position and spectral
dispersion, respectively. This can be written as a matrix relation:
\begin{equation}
\left(
\begin{array}{cc}
X\\
Y
\end{array}
\right)  =  \left(
\begin{array}{cc}
a(x,\lambda) & b(x,\lambda) \\
c(x,\lambda) & d(x,\lambda)
\end{array}
\right)
\left(
\begin{array}{cc}
x\\
\lambda
\end{array}
\right)  =  \left(
A
\right)(x,\lambda)
\left(
\begin{array}{cc}
x\\
\lambda
\end{array}
\right)
\end{equation}

What we need then is to invert the matrix $(A)(x,\lambda) = \left(
\begin{array}{cc}
a(x,\lambda) & b(x,\lambda) \\
c(x,\lambda) & d(x,\lambda)
\end{array}
\right)$ to get the distortion relation:

\begin{equation}
\left(
\begin{array}{cc}
x\\
\lambda
\end{array}
\right)  =  (A)^{-1}(X,Y)
\left(
\begin{array}{cc}
X\\
Y
\end{array}
\right)
\end{equation}

{\it Spatial Direction:}

The spatial grid is first used. The $a$ parameter is estimated
by fitting a polynomial to the position of the different features
detected on the beams (3 features per beam):

\begin{equation}
a(x,\lambda) = a_0(\lambda) + a_1(\lambda) \times x + a_2(\lambda) \times x^2 + ...
\end{equation}

{\it Spectral direction:}

We are then only interested in determining the $\lambda$-dependent
distortion. Therefore, a ``normal'' $\lambda$-calibration frame
(i.e. with the set of absorption plastic foils, or using sky
absorption lines) should be available for that. In that case, one can
determine the $d$ coefficient:

\begin{equation}
d(x,\lambda) = d_0(x) + a_1(x) \times \lambda + d_2(x) \times \lambda^2 + ...
\end{equation}

We note here that this first step will give a coarse
wavelength-calibration, which will be refined in a further step using
telluric lines.

\subsubsection{Shift and zoom, $\kappa$-matrix} 

As shown in Fig.~\ref{fig:fringes} and Fig.~\ref{fig:shift_map}, the
MATISSE photometric beams have a different width than the
interferometric beam, for obvious pixels-saving reasons. Therefore,
since the spatial filtering is not perfect (due to the fact that we
use pinholes instead of fibers, for technology readiness reasons), one
needs to scale the photometric beams to the interferometric beam size
by shifting and zooming them to the right place and scale.

In addition, one has to get the flux ratio coefficient between the
flux measured in the interferometric beam and the one measured in the
photometric beam, in order to have all the photometric information
available to compute the observables.

\begin{figure}[htbp]
\begin{center}
\begin{tabular}{cccc}
\includegraphics[width=0.22\textwidth]{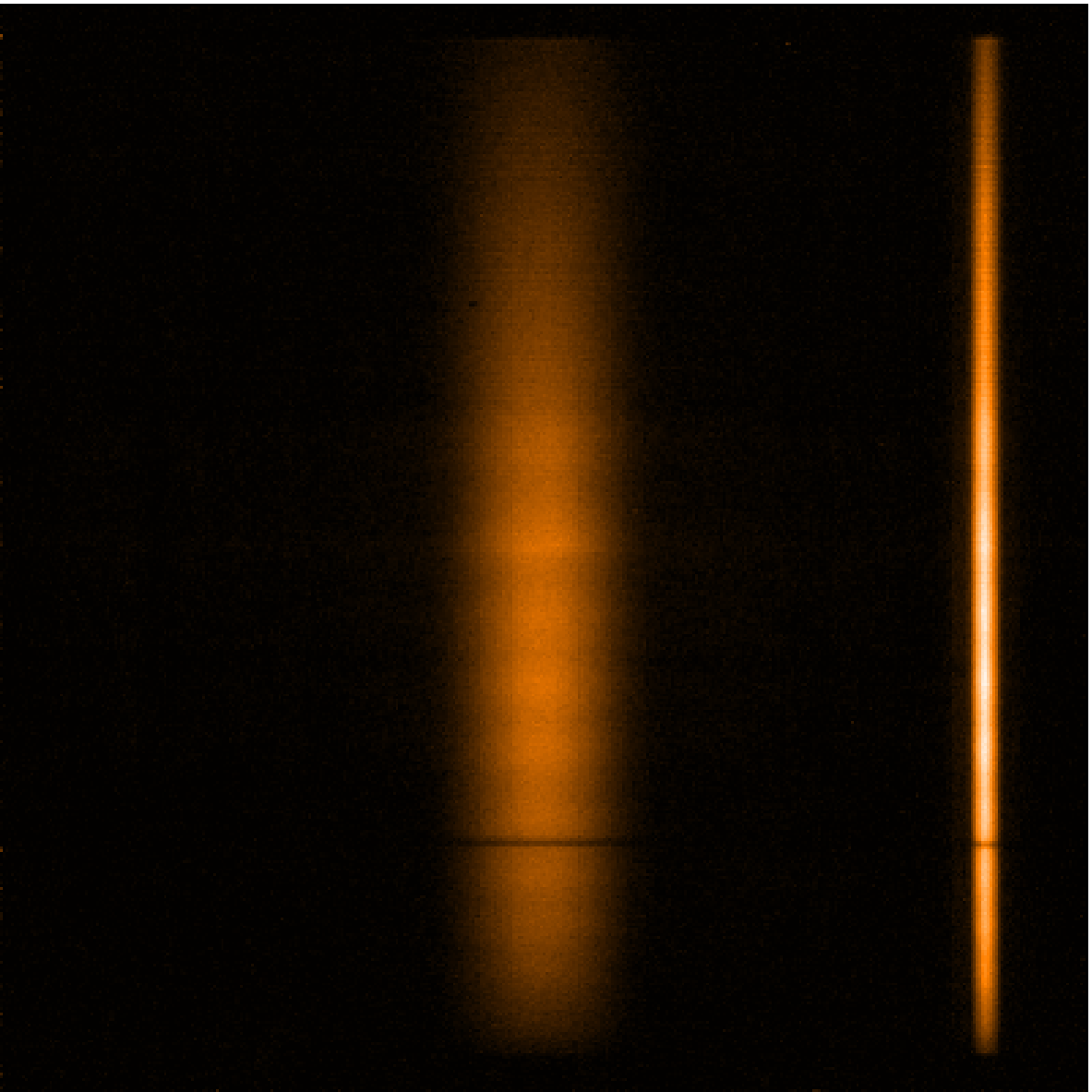}&
\includegraphics[width=0.22\textwidth]{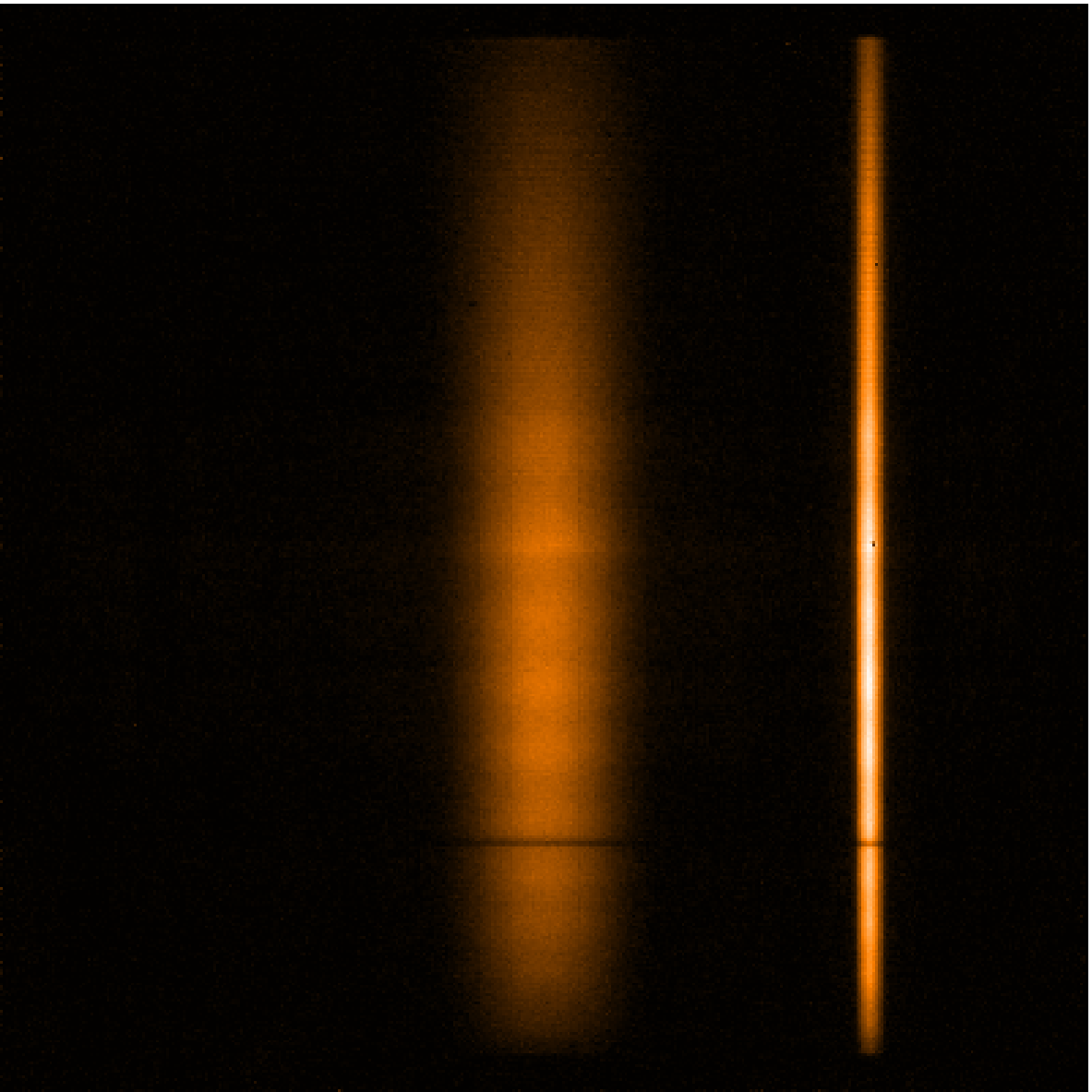}&
\includegraphics[width=0.22\textwidth]{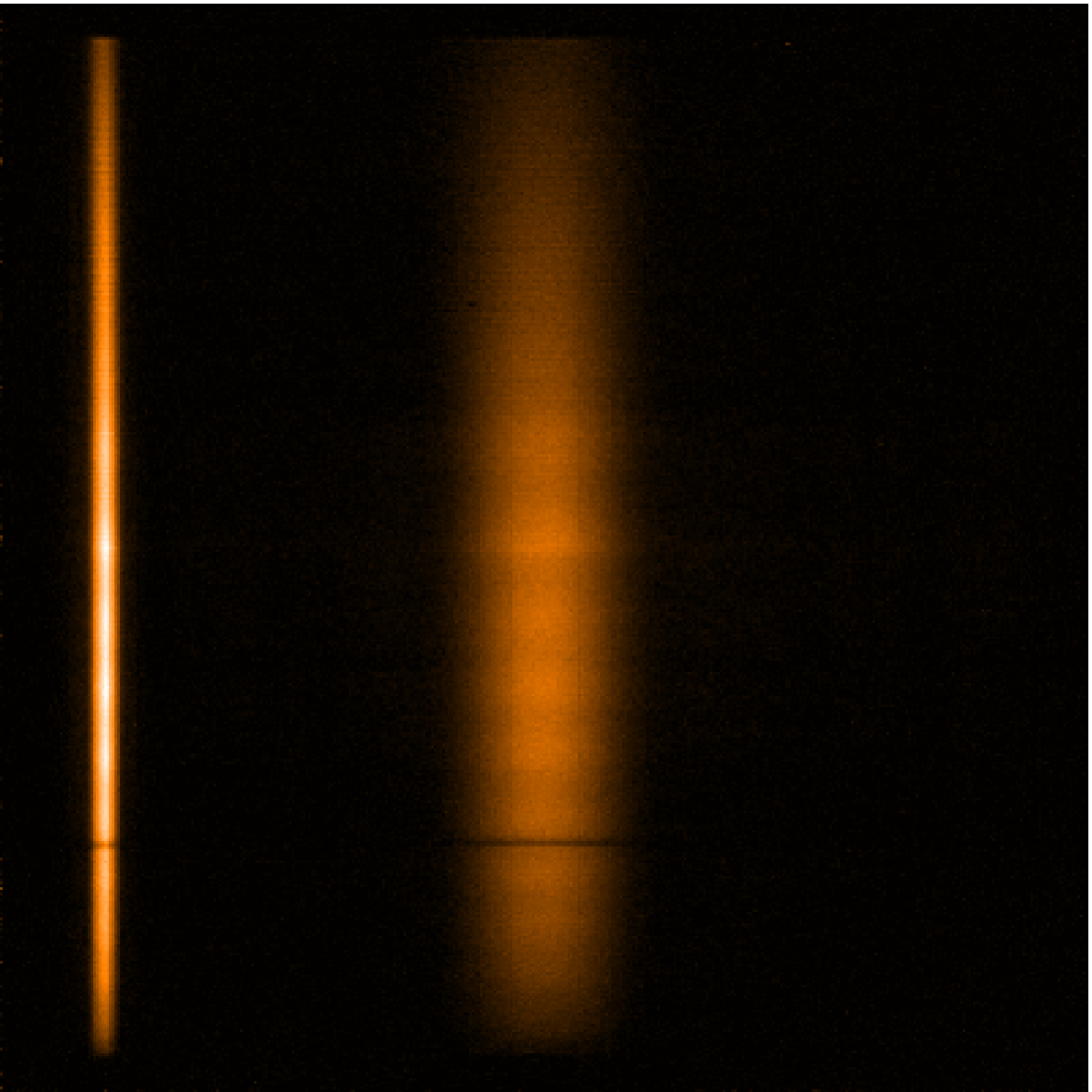}&
\includegraphics[width=0.22\textwidth]{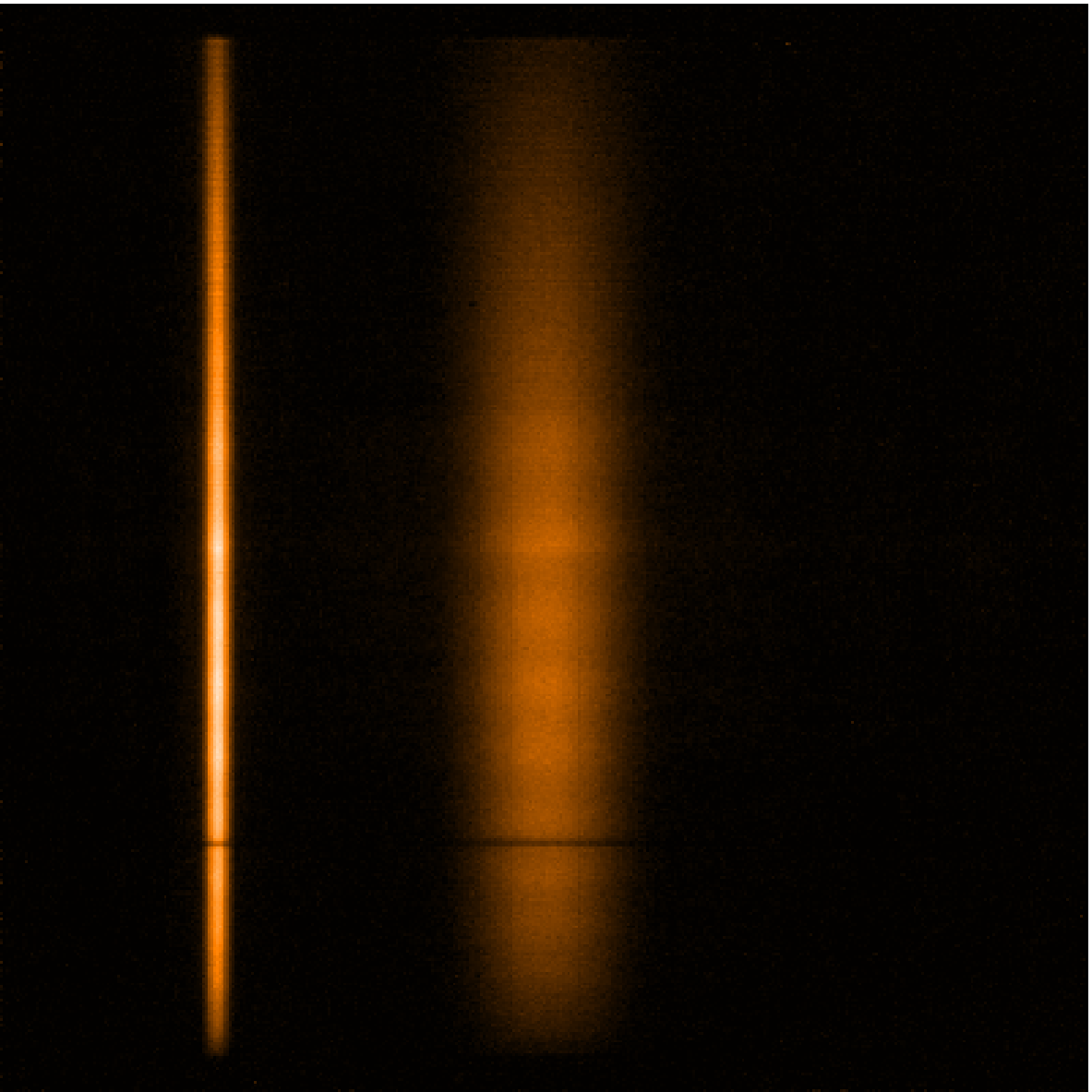}
\end{tabular}
\end{center}
\caption[Kappa Matrix] 
{ \label{fig:kappa} Kappa matrix acquisition sequence in
the N-band, with alternatively just one out of four open
shutter.  }
\end{figure}

The Shift and Zoom coefficients, together with the $\kappa$-matrix are
computed using 4 illuminated frames, one for each of the telescope
beams, the other beams being closed by shutters (see
Fig.~\ref{fig:kappa}). The source is preferably artificial, part of
MATISSE instrument, or could be a bright unresolved astronomical
target as well.

The photometric contribution $P'_i(x, \lambda)$ of the telescope $i$ to
the interferometric beam can be expressed as following:
\begin{equation}
P'_i(x, \lambda,t) = \kappa_{ij}[x', \lambda, s_j(\lambda),
z_j(\lambda)] \times P_{ij}(x', \lambda,t) \label{eq:KappaMatrix}
\end{equation}

where $P_{ij}(x', \lambda,t)$ is a vector containing the contribution
of the telescope $i$ to the photometric beams $j$ (after chopping
correction, see section \ref{sec:chopping}), $\kappa_{ij}$ is the
linear transformation matrix of the intensities in the photometric
channels into the interferometric channel (the so-called
"$\kappa$-matrix"), and $s_j(\lambda)$, $z_j(\lambda)$ are the shift
offset and zoom coefficient to match the photometric beam $j$ into the
interferometric beam.

We determine therefore the $\kappa$-matrix and the shift-and-zoom
coefficients by fitting the photometric beam shape and intensity to
the interferometric beam for each selected wavelength.

\subsubsection{Applying cosmetics}
\label{ssec:cosmetics}

The first steps of the data reduction are:
\begin{enumerate}
\item Subtracting the average (cold) dark ${\rm DARK}(X,Y)$ in each
frame,
\item Compensating of the space-variant gain in each frame by division
of each frame through the ${\rm FFM}(X,Y)$ (flat-field map),
\item Interpolating detector bad pixels in each frame with the ${\rm
BPM}(X,Y)$ (bad pixel map),
\item Applying the distortion matrix to transform $(X,Y)$ to $(x,\lambda)$,
the coordinates of the data,
\item Adjusting the photometric contribution in the interferometric
beam by applying the $\kappa$-matrix, shift, and zoom coefficients
\end{enumerate}

\begin{figure}[htbp]
\begin{center}
\begin{tabular}{cccc}
\includegraphics[height=0.25\textwidth]{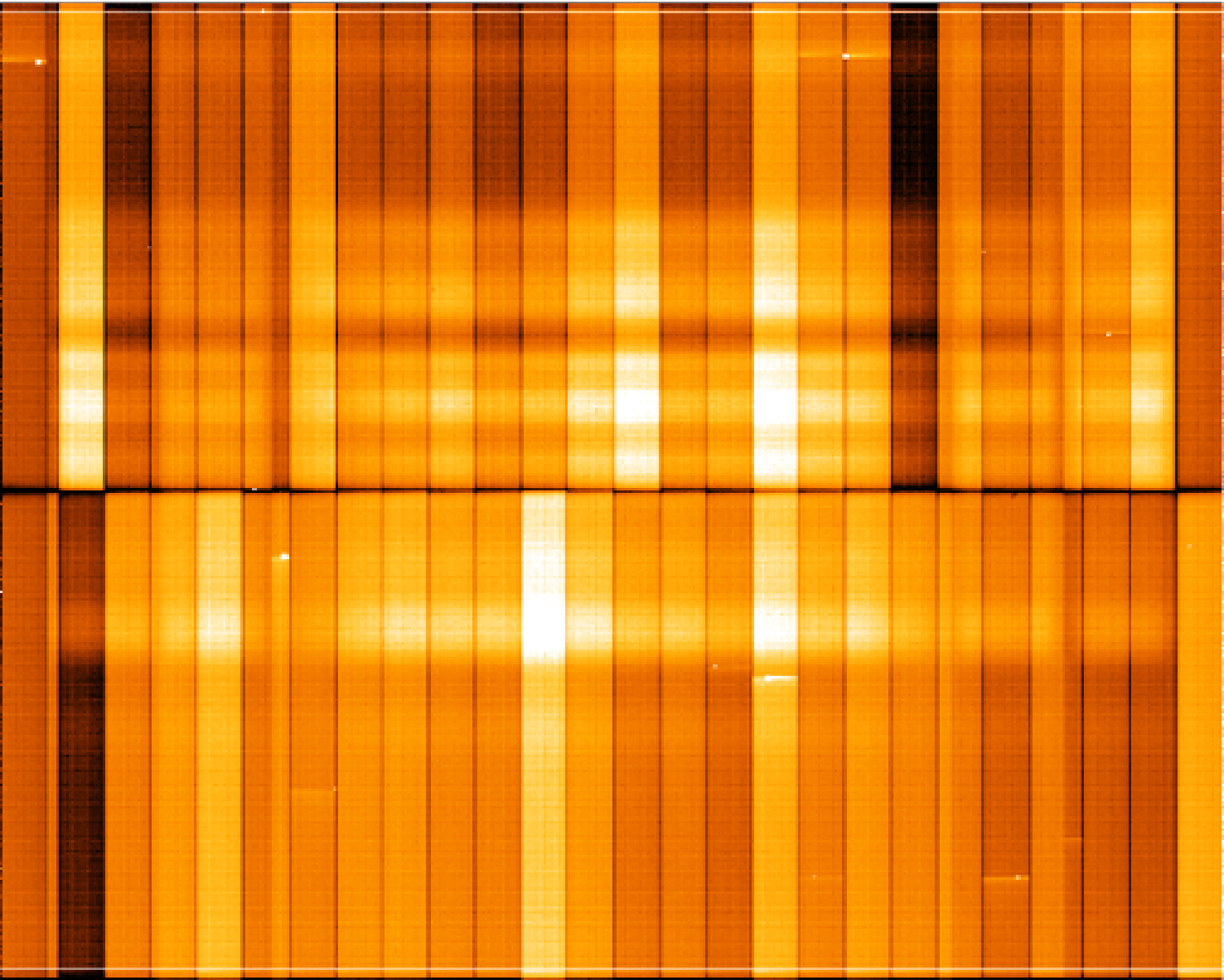}&
\includegraphics[height=0.25\textwidth]{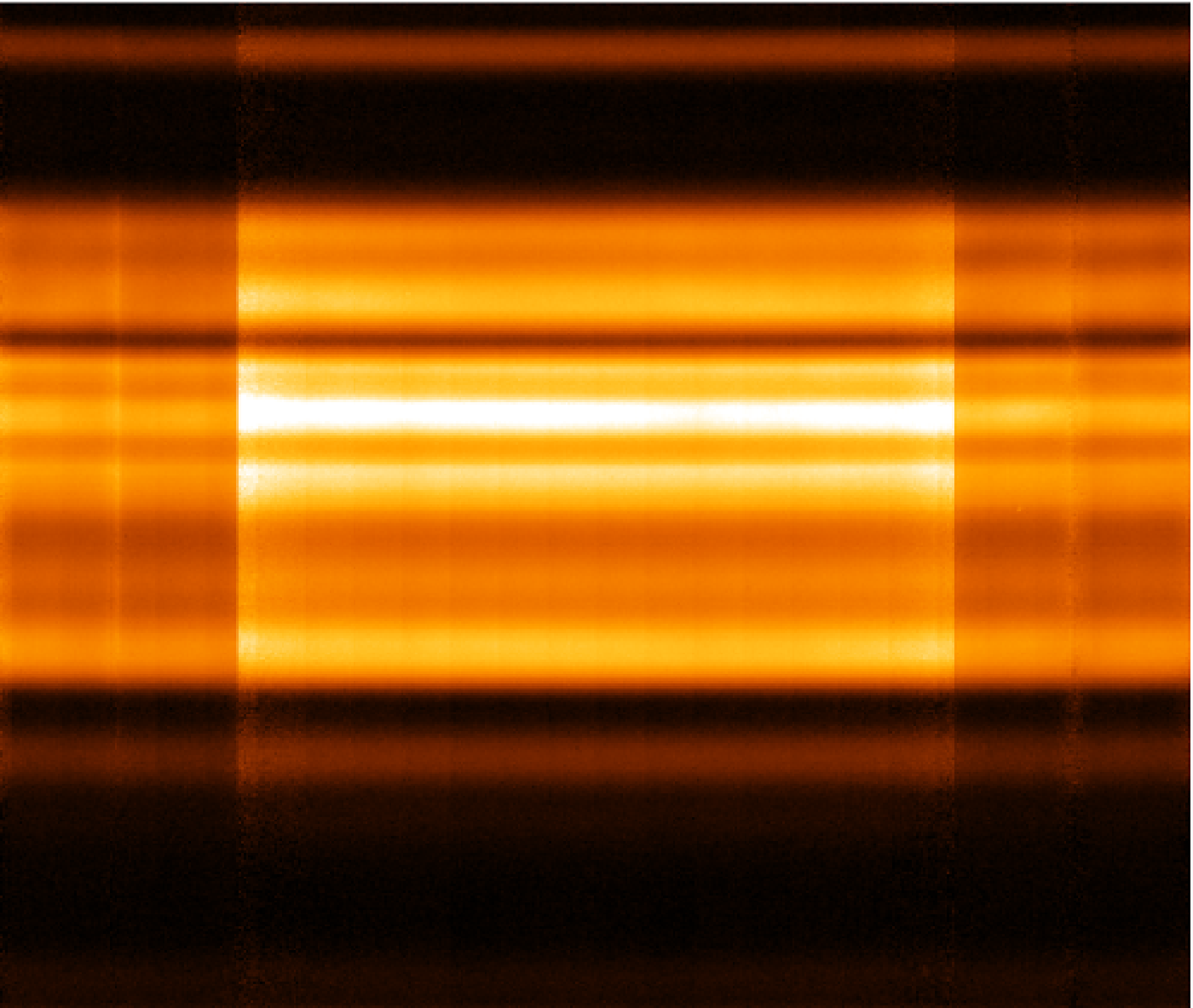}
\end{tabular}
\end{center}
\caption[Distortion Estimation] 
{ \label{fig:flat_field}
{\bf Left:} The raw image from the Aquarius detector (7-13 $\mu$m) contains four photometric channels,
the interferometric channel and 16 reference areas, showing bad pixels and detector channel
offsets.
Right: The same image after the calibration step contains the photometric channels and the
interferometric channel only.}
\end{figure}

The result of treating the three first effects is shown in
Fig.~\ref{fig:flat_field}. The two last corrections are being tested
right now on the mounted instrument in the Nice laboratory.

The results of these processes are: a cleaned-up fringe pattern
$i^{'}(x, \lambda, t)$, and cleaned-up photometric estimates
$P^{'}_i(x, \lambda, t)$.

\subsection{How to tackle the multiple-stage modulations in MATISSE} 

As described in Petrov et al. (2016)\cite{Petrov2016}, MATISSE
includes a three-stage modulation process for the mitigation of the
sky background (as MATISSE works in the mid-infrared). These stages
are namely chopping, spatial modulation and temporal modulation, in
addition to beam commutation, which is used to clean up the phases out
of the instrumental effects.

\subsubsection{Chopping}\label{sec:chopping}

Chopping produces alternatively science and sky frames with a
switching frequency as fast as 1Hz. The first step is to remove these
sky frames to the science frames, resulting in removing the bulk of
the background contribution ($\approx99$\%):

\begin{equation}
i(x, \lambda, t) = i^{\rm \,obj}(x, \lambda, t) - i^{\rm \,sky}(x, \lambda, t) 
\end{equation}

\subsubsection{Spatial modulation} 

Spatial modulation separates, in the Fourier plane, the fringes energy
(``fringe peak'') from the flux energy (``photometric
peak''). Therefore, a simple Fourier processing along the $x$ axis will cancel further
out the contribution of the background to the data.
Correlated fluxes are computed just by Fourier-Transforming the clean
interferogram:
\begin{eqnarray}
\nonumber
I^{'}(u, \lambda, t) & = & {\rm FT}_x[i(x, \lambda, t)]\\ 
& = & F_{00}(u,\lambda,t) + \sum_{i=1}^4 \sum_{j=i+1}^4 \left[
F_{ij}(u-f_{ij}, \lambda, t) \cdot V_{ij}(\lambda) \cdot
e^{\i \Phi_{ij}(\lambda,t)} \right]
\end{eqnarray}

where $u$ is a spatial frequency, $f_{ij}$ is the fringe rate, as
explained below eq.~\ref{eq:interferogram}, $V_{ij}(\lambda)$ is the
target visibility, $\Phi_{ij}(\lambda,t)$ is the phase of the target
plus piston, and $F_{ij}(u,\lambda,t)$ is a function of the
photometric contributions $P'_i(x, \lambda)$, involving Fourier
Transforms and correlations, representing the shape of the peaks. We
note that $\sum\limits_{u}F_{ij}(u, \lambda, t)
= \sqrt{\sum\limits_{x} P_i(x,\lambda,t) \cdot P_j(x,\lambda,t)}$, and
that $F_{00}(u,\lambda,t)$ represents the photometry peak.

The background residuals left after chopping contribute only to
$F_{00}(u,\lambda,t)$ at first order, as the background does not
contain any coherent information. At the frequency of each fringe peak
$f_{ij} = b_{ij}/\lambda$, we have:

\begin{equation}
I^{'}_{ij}(\lambda, t)  = I^{'}(f_{ij}, \lambda, t) = 
F_{ij}(0,\lambda,t) \cdot V_{ij}(\lambda) \cdot e^{\i\Phi_{ij}(\lambda,t)} +
B^{\rm res}_{ij}(\lambda,t)
\end{equation}

where the contribution of the residual background $B^{\rm
res}_{ij}(\lambda,t) = F_{0}(f_{ij},\lambda,t)$ is much smaller than
at the frequency 0 (as seen in Fig.~\ref{fig:fringepeaks} on top and
bottom row).

\subsubsection{Temporal modulation}

In addition to the spatial modulation, a temporal modulation is
introduced by piezoelectric actuators, changing the optical path
difference (OPD) $\Delta^{\rm mod}_{ij}(t)$ by a fraction of the
wavelength between consecutive frames. Let us consider that there are
$N_T$ steps over an OPD range of $\Delta_{ij}$ during one modulation
cycle $T$. The modulation is then $\Delta^{\rm mod}_{ij}(t)
= \frac{k}{N_T}\Delta_{ij} $, $k$ being the modulation step
number. 


On needs to consider the fringes ``freezed'' over the modulation time
$T$ in order to consider $\Delta^{\rm atm}_{ij}(t)$
constant. Therefore, the modulation cycle must be done over one
atmosphere coherence time $T\approx\tau_0$, or in conjunction with a
fringe tracker which freezes the fringes over a time
$T\approx\tau_{\rm FT}$.

A de-modulation is applied to the previous Fourier-Transform,
baseline-by-baseline, and we integrate the fringe peak
$I^{'}_{ij}(\lambda, t)$ over one modulation cycle (the $N_T$ steps of
modulation). We do so by multiplying it with a phasor containing the
counter-modulation:

\begin{eqnarray}
\nonumber
I^{''}_{ij}(\lambda, T) &=& \frac{1}{N_T} \sum_{k=0}^{N_T}
I^{'}_{ij}(\lambda, t) \times e^{-2i\pi \Delta^{\rm mod}_{ij}(t)
/ \lambda}\\
&=& F_{ij}(0,\lambda,t) \cdot V_{ij}(\lambda) \cdot
e^{i\Phi_{ij}(\lambda,t)} +
\delta B^{\rm res}_{ij}(\lambda,t)
\label{eq:coherentFlux}
\end{eqnarray}

The background residual is further reduced by the modulation, leading
to the final residual background $\delta B^{\rm res}_{ij}(\lambda,t)$:

\begin{equation}
\delta B^{\rm res}_{ij}(\lambda,t) = B^{\rm res}_{ij}(\lambda,t) \times \frac{1}{N_T}\sum_{k=0}^{N_T} e^{\left[-2 i \pi k \cdot \Delta_{ij} / (N_T\cdot \lambda) \right]} 
\label{eqBres}
\end{equation}

\begin{figure}[htbp]
\begin{center}
\begin{tabular}{cccc}
\includegraphics[width=0.3\textwidth]{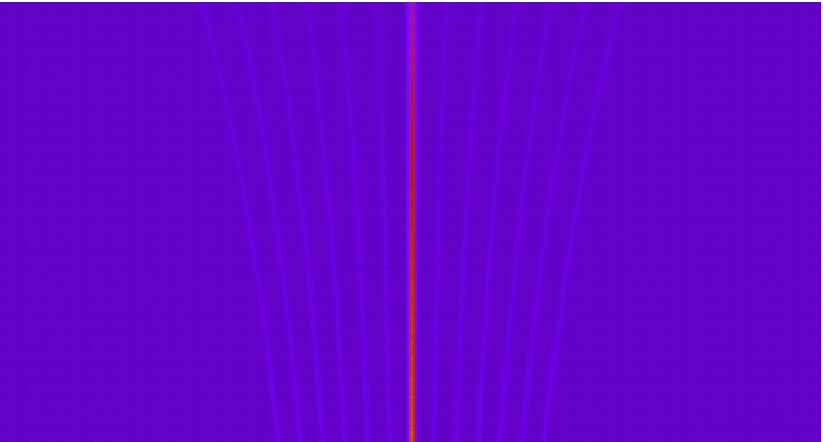}&
\includegraphics[width=0.3\textwidth]{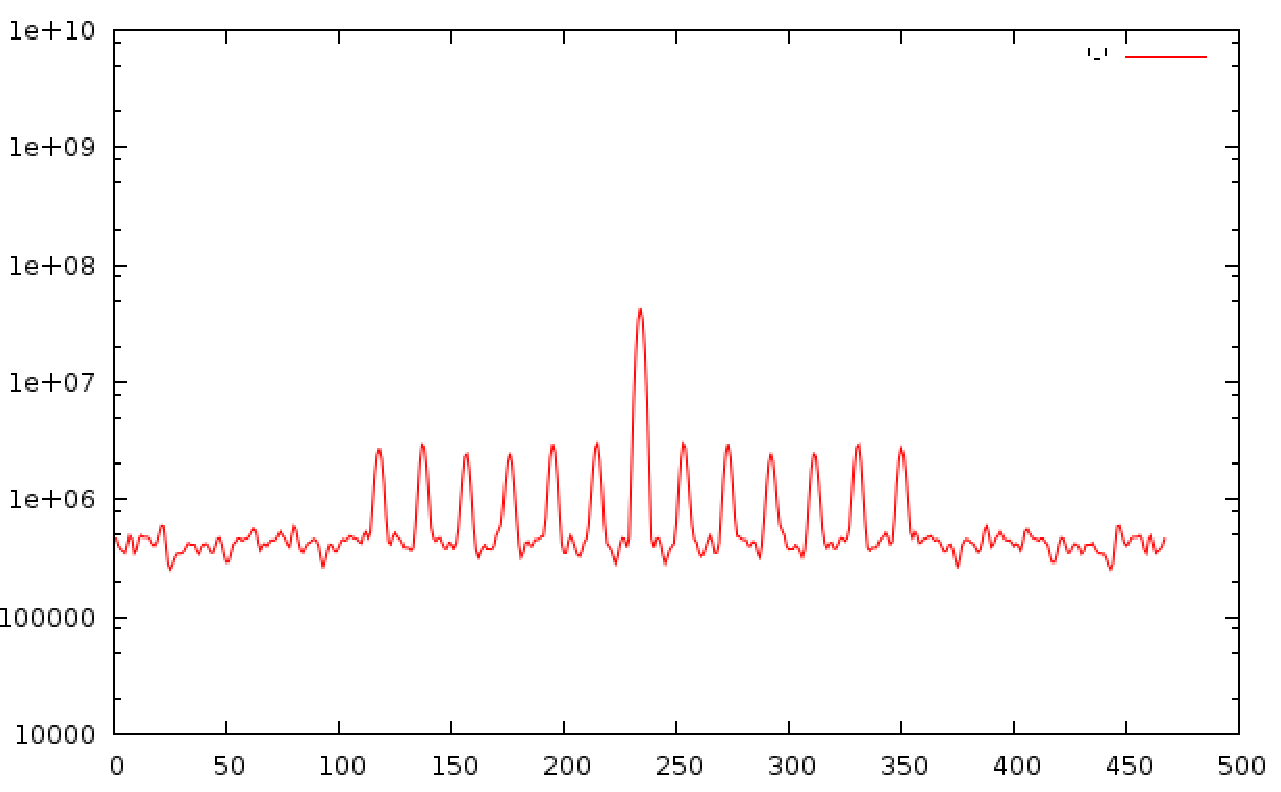}\\
\includegraphics[width=0.3\textwidth]{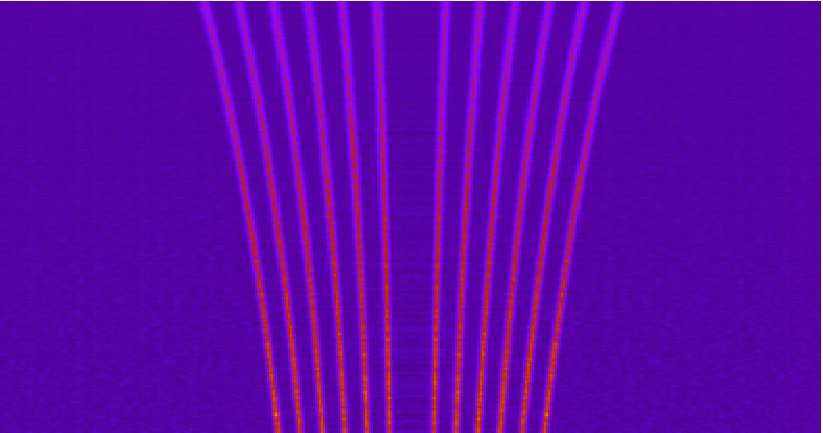}&
\includegraphics[width=0.3\textwidth]{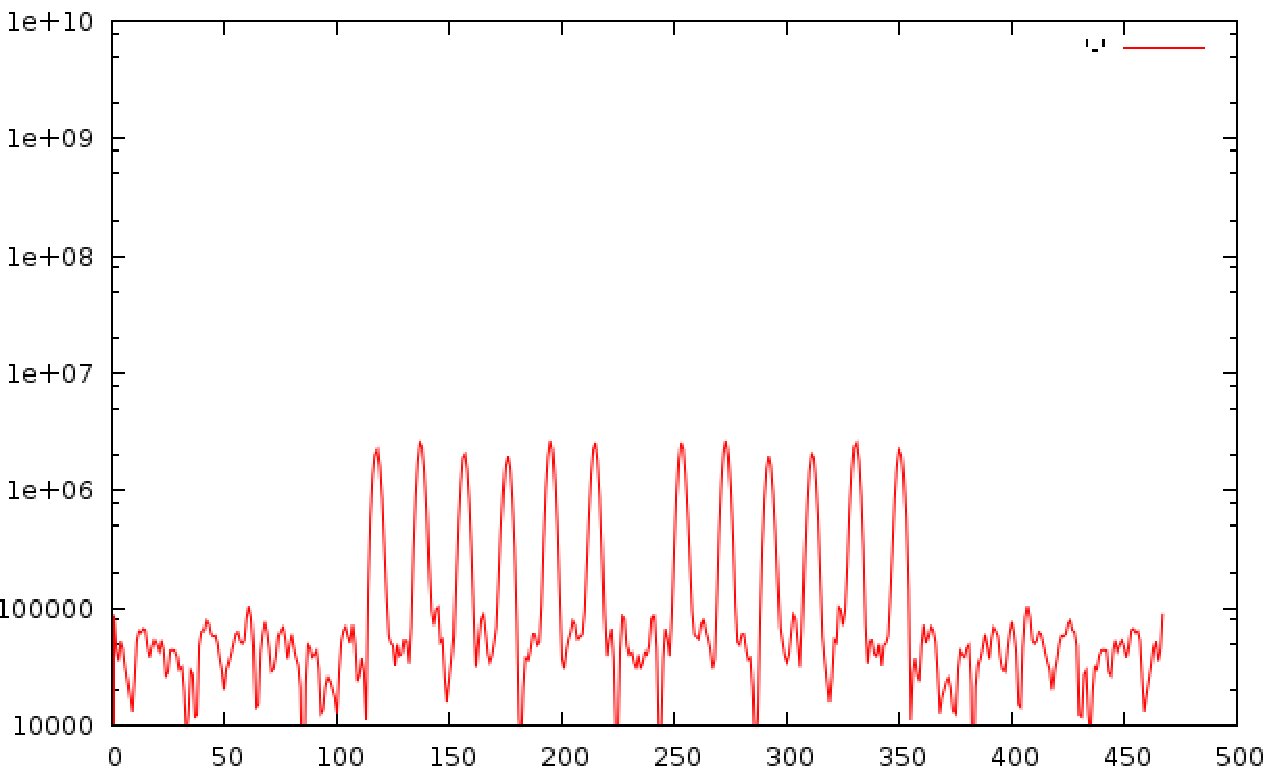}\\
\includegraphics[width=0.25\textwidth]{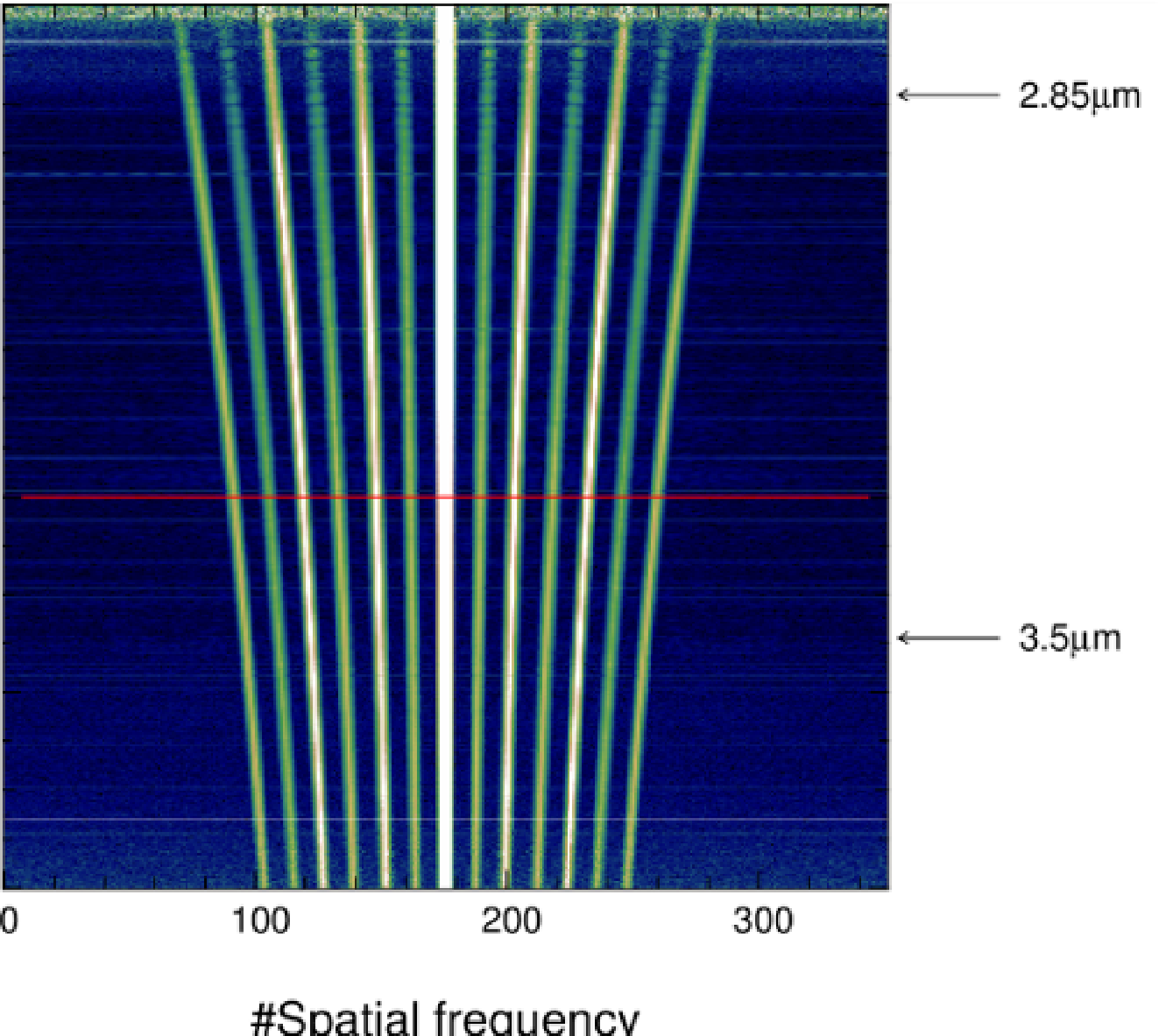}&
\includegraphics[width=0.3\textwidth]{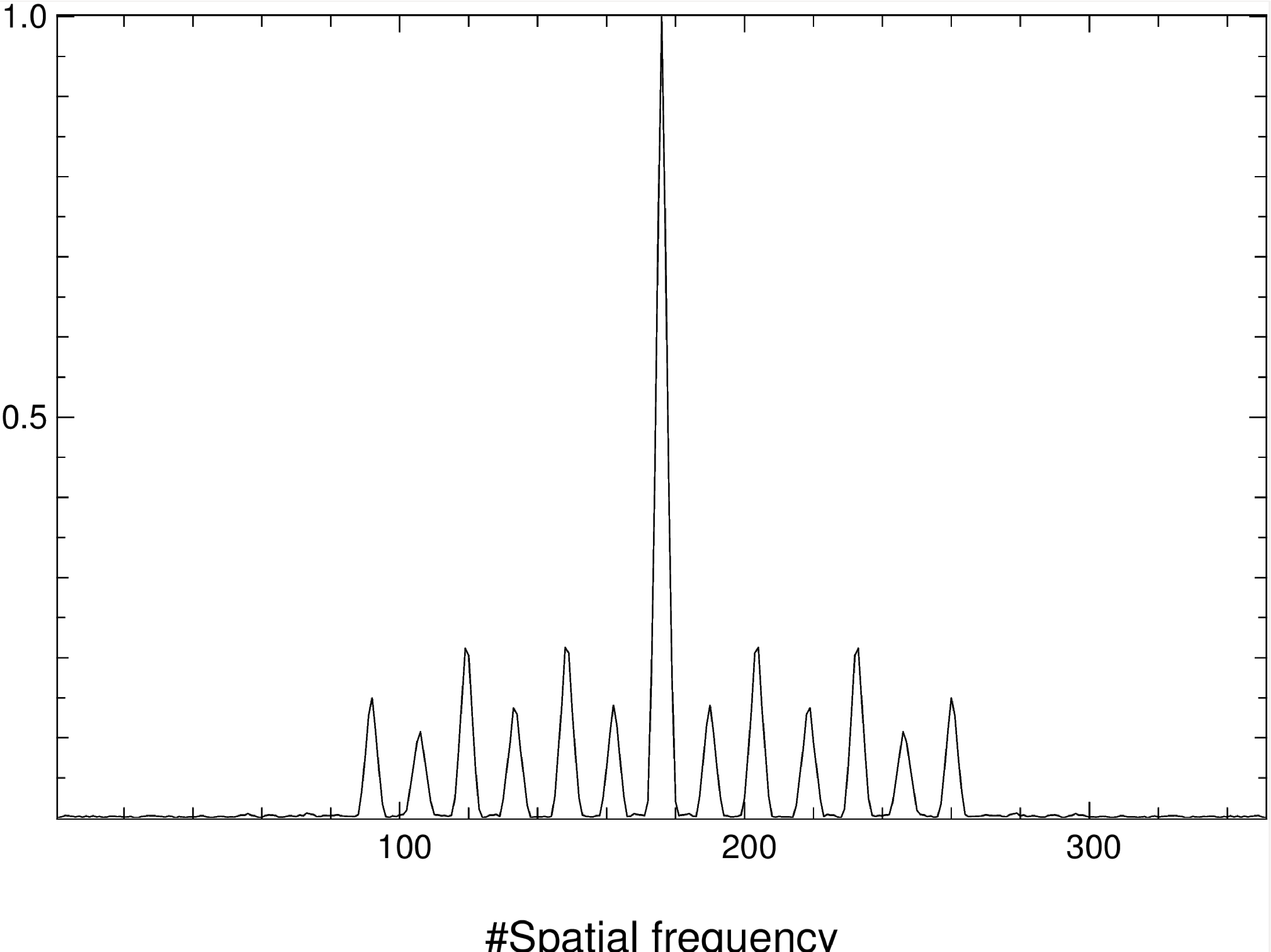}\\
\end{tabular}
\end{center}
\caption[Kappa Matrix] 
{ \label{fig:fringepeaks} {\bf Left column:} Fourier
transform of the MATISSE fringe pattern.
{\bf Right column:} cut at one wavelength. {\bf Top:}
Simulated fringe pattern after the chopping
correction. {\bf Middle:} Same as before, but after the
temporal demodulation applied. {\bf Bottom:} Same as first
one but on the real first fringes data shown in
Fig.~\ref{fig:fringes}. The (de)modulation is not yet
operational on the instrument.}
\end{figure}

The effect of demodulation can be seen in Fig.~\ref{fig:fringepeaks},
middle row. The combination of spatial modulation, temporal modulation
and chopping are the three ways selected in MATISSE to tackle the
dominance of the background. Please note that neither the spatial nor
the temporal modulation can be applied to the photometries. Therefore,
the rejection of the sky background in the photometric estimates rely
only on Chopping, and will be of much lower quality.

\subsection{Estimators} 

Once the complex coherent flux $I^{''}_{ij}(\lambda, T)$ has been
computed, it is possible to compute the estimators, i.e. the raw
visibilities and phases of the object. These estimators are split
between \emph{incoherent estimators} (speckle-like), that can be
computed without the knowledge of the optical path difference, and
\emph{coherent estimators} that need an appropriate atmospheric OPD
estimate, be it simply due to the turbulence or to higher-order
effects (longitudinal dispersion).

\subsubsection{``Incoherent'' estimators}

\paragraph{Squared visibility}

Squared visibilities are computed as follows:

\begin{equation}
V_{ij}^2 (\lambda) = 
\frac{\sum\limits_{u} \left<|I^{'}(u,\lambda, t)|^2 - \beta \right>_t}
{2\cdot \sum\limits_{x} \left<\overline{P_a}'(x,\lambda,t)\cdot \overline{P_b}'(x,\lambda,t)\right>_t}
\label{eq:subsec3.8-5}
\end{equation}

where $u$ is the frequency around the considered baseline, and $\beta$
a bias estimated on $I^{'}$ outside the range of frequencies where the
fringe peaks are present, and $\overline{P_a}'(x,\lambda,t)$ and
$\overline{P_b}'(x,\lambda,t)$ are the estimates of the photometric
fluxes transformed into the interferometric channel (see
eq.~\ref{eq:KappaMatrix}).

In the case where there are no photometries recorded (like in
HIGH-SENS mode without a photometric acquisition sequence), the
squared visibilities are simply not computed. Instead, the user should
use the correlated fluxes, i.e. the upper part of
eq.~\ref{eq:subsec3.8-5}:

\begin{equation}
{\rm CF}_{ij}^2 (\lambda) = 
\sum\limits_{u} \left<|I^{'}(u,\lambda, t)|^2 - \beta \right>_t
\label{eq:corrFlux}
\end{equation}

\paragraph{Closure Phase} 

\begin{figure}[htbp]
\begin{center}
\begin{tabular}{c}
\includegraphics[angle=-90]{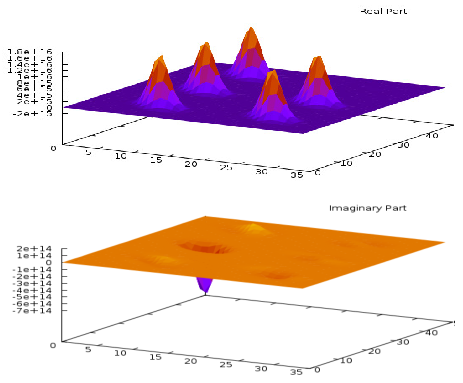}
\end{tabular}
\end{center}
\caption[Kappa Matrix] 
{ \label{fig:bispectrum} Example bispectrum values computed
on simulated MATISSE data. Top graph: real part, and
bottom graph: imaginary part. The middle-left peak is
spurious. }
\end{figure}

The bispectrum is calculated by:

\begin{equation}
O^{(3)}_{ij} (\lambda) = 
\left<
I^{'}(u,\lambda, t) \times
I^{'}(v,\lambda, t) \times
{I^{'}}^{*}(u+v,\lambda, t) \right>_t
\end{equation}

where $u$, $v$, and $u+v$ are the frequencies of the baselines
involved in the closure relation. An example MATISSE bispectrum can be
seen in Fig.~\ref{fig:bispectrum}. Out of the 5 peaks that can be
seen, just 4 contain useful information from the 4 closure phases
available at 4 telescopes (the middle-left peak is a spurious peak
coming from the contamination of the central photometric peak).

This average bispectrum contains an additive noise bias terms plus a
multiplicative one, because of the all-in-one combination of the
interferograms.  These biases need to be subtracted in order to get
closure phases without systematic errors, following specific
recipes\cite{1985JOSAA...2...14W, 2012A&A...541A..46G}.

\subsubsection{``Coherent'' estimators}

Coherent estimators need an appropriate OPD estimate. We describe here
briefly what we envision to include in \texttt{drsmat}. The complex
coherent flux is provided by the equation~\ref{eq:coherentFlux}.

The phase term $\Phi_{ij}(\lambda, t)$ depends on the variations of
the atmospheric properties (temperature, pressure), producing an
achromatic variation of OPD (but not phase), and it depends also on
the refractive index $n(\lambda))$ of air, which is chromatic and
dependent of its water vapor composition. It can be expressed as:

\begin{equation}
\Phi_{ij}(\lambda, t) = \Phi_{ij}^{\rm obj}(\lambda) + 2\pi
\frac{ \Delta_{ij}^{\rm atm}(t) }{ \lambda} +
2\pi \frac{\Delta_{ij}^{\rm atm}(t) }{ \lambda } \cdot [n(\lambda,
t) - 1]
\end{equation}

the different terms are the following:

\begin{itemize} 
\item $\Delta_{ij}^{\rm atm}(t)$: achromatic part of
the atmospheric OPD, 
\item $[n(\lambda, t)-1] \approx a(\lambda) +
b(t) $, chromatic term with $a(\lambda)$ wavelength-dependent but
roughly static \cite{Ciddor1996, Mathar2007}, and $b(t)$ wavelength
independent, but with strong variations in
time \cite{2004SPIE.5491..715J} 
\item $\Phi_{ij}^{\rm obj}(\lambda)
= \Phi_{ij}^{(0)} + \Phi_{ij}^{(1)} / \lambda + \Phi_{ij}^{\rm
diff}(\lambda)$: the object phase, which can be decomposed in a
Taylor series containing the differential phase $\Phi_{ij}^{\rm
diff}(\lambda)$.  
\end{itemize}

\paragraph{Achromatic Optical Path Difference estimates:}
\label{section3:subsection7}

MATISSE will compute the achromatic OPD using a VEGA-like
algorithm\cite{2009A&A...508.1073M, 2011A&A...531A.110M}: a 2D-Fourier
Transform (FT) is computed, to take profit of the multi-axial beam
combination of MATISSE. The fringe peak in the power spectrum appears
as an offset peak, whose offset is proportional to the OPD. The great
advantage of this method is that the peak is concentrated onto a small
location of the 2D map, whereas the noise is spread everywhere, hence
contributing to a smaller fraction of the peak noise. In addition,
there is no ambiguity at OPD 0, contrary to a MIDI-like estimate.

This 2D-FT can be computed in two steps, first in the spatial
direction and second in the spectral direction, allowing us to
counteract the OPD modulation pattern before elevating the Fourier
Transform to its square, and hence increasing tremendously the SNR of
the peak detection. This is also the way the fringe pattern will be
detected in real time with the instrument.

Another option implemented for OPD determination is a fit to the
complex coherent flux phase, in the very same way as it is done for
AMBER \cite{2007A&A...464...29T}.

The computed atmospheric OPD contains then the two terms: 
${\rm OPD} = \Delta_{ij}^{\rm atm}(t) + \Phi_{ij}^{(1)}$.

\paragraph{Chromatic OPD estimates:}

ESO provides information through the ambient conditions monitor that
will be sufficient to correct the data from the chromatic dry and wet
air dispersion, down to accuracies of 3 degrees in K band and down to
10 degrees in N-band.

The static water vapor term $a(\lambda)$ will be computed using the
pressure $P$, temperature $T$ (in the tunnels), and relative humidity
of the ambient air $R_h$. We add to these the partial pressure of
CO$_2$, which can be hard-coded to an average value at Paranal. The
fraction of humidity is converted into partial pressure of water vapor
$p_{wv} = R_h \times P$.  The $\lambda$-dependent shape of the
chromatic phase $a(\lambda)$ is then be determined based on computed
optical index of air $n$ as a function of wavelength \cite{Ciddor1996,
Mathar2007}.  The next step is to compute phases, and therefore we
also use the information of the delay lines position to compute the
difference of delay for each baseline: if $A_i$ is the static optical
path introduced by the delay line $i$, and $OPL_i$ the dynamic path,
$A_i+OPL_i$ is the total path of air introduced by each delay
line. Per baseline $ij$, the path to use for the chromatic OPD is
therefore $\delta_{ij} = A_i+OPL_i -A_{j}-OPL_{j}$. The chromatic
phase is then computed by $\phi^{\rm chrom}(\lambda) = 2 \pi
a(\lambda)
\frac{\delta_{ij}}{\lambda}$, with $n$ the computed index of air,
and $\lambda$ the wavelength of interest.

The water vapor phase offset, time-variable, $\Delta_{ij}^{\rm
atm}(t) \cdot b(t)$, will be obtained by averaging the phase term in
the wavelength-direction after subtraction of the atmospheric
OPD. As a consequence, it also removes the object-related phase
offset $\Phi_{ij}^{(0)}$. The result is therefore $\Delta_{ij}^{\rm
atm}(t)\cdot b(t) + \Phi_{ij}^{(0)}$.

All these calculations yield $\Delta_{ij}^{\rm atm}(t) + \Phi_{ij}^{(1)}$,
$a(\lambda)$, and $\Delta_{ij}^{\rm atm}(t) \cdot b(t) + \Phi_{ij}^{(0)}$.

\paragraph{Differential phase:} 

After that step, the computed phase is transformed into a differential
phase the standard way \cite{2006EAS....22..379M,2012ASPC..464...15M,
2014EAS....69...17M}. We recall it briefly here. First, we multiply
the Fourier transform of the correlated interferograms by the counter
OPD phasor. This yields the OPD-free complex coherent flux:

\begin{eqnarray}
\nonumber
{\rm CF}_{ij}(\lambda, t) & = & I^{''}_{ij}(\lambda, t) \times
e^{-2i\pi\,\left( \left[ \Delta_{ij}^{\rm atm}(t) / \lambda
+ \Phi_{ij}^{(1)}\right] + \left[ \Delta_{ij}^{\rm atm}(t)\cdot
b(t)+\Phi_{ij}^{(0)} \right] + a(\lambda) \right) } \\ 
& = & F_{ij}(0,\lambda,t) \cdot V_{ij}(\lambda) \cdot 
e^{i\, \Phi_{ij}^{\rm diff}(\lambda)}
\end{eqnarray}

Averaging this OPD-free complex coherent flux over time yields the
following products:

\begin{itemize}
\item $<F_{ij}(0,\lambda,t)>\cdot V_{ij}(\lambda)$ : correlated flux degraded
by the not corrected beam overlap
\item  $\Phi_{ij}^{\rm diff}(\lambda)$ : differential phase, with any
wavelength-linear phase term set to zero.
\end{itemize}

\paragraph{Coherent visibility:}

Finally, it is possible to continue the processing to lead to a linear
visibility estimate. For that, we divide the correlated flux by the
photometric factor. This is the last optional step, yielding the
linear visibility estimate:

\begin{equation}
V_{ij}(\lambda)) = \left| \frac{ \sum \limits_{u} \left< {\rm
CF_{ij}(u, \lambda, t)} \right>_t }{\sqrt{ \left< \sum\limits_{x}
P_i(x,\lambda,t) \cdot P_j(x,\lambda,t)\right>_t} } \right|
\end{equation}

In the case where there are no photometries recorded (like in
HIGH-SENS mode), this term is not computed, as for the squared
visibilities.

\subsection{Data calibration} 

To calibrate the data, we apply the same recipes as in
AMBER\cite{2008SPIE.7013E.132M}, i.e.:
\begin{itemize}
\item We start with searching calibrator stars diameters in published catalogs,
\item we divide the calibrators visibilities by their expected visibilities and store the transfer function result in new files,
\item we interpolate the transfer function to the time of the science observations with a Gaussian-weighted average (typical FWHM 1hr),
\item we divide the science visibilities with the interpolated
transfer function and store the result as calibrated science-grade
data files.
\end{itemize}

\begin{figure}[htbp]
\begin{center}
\begin{tabular}{ccc}
\includegraphics[width=0.3\textwidth]{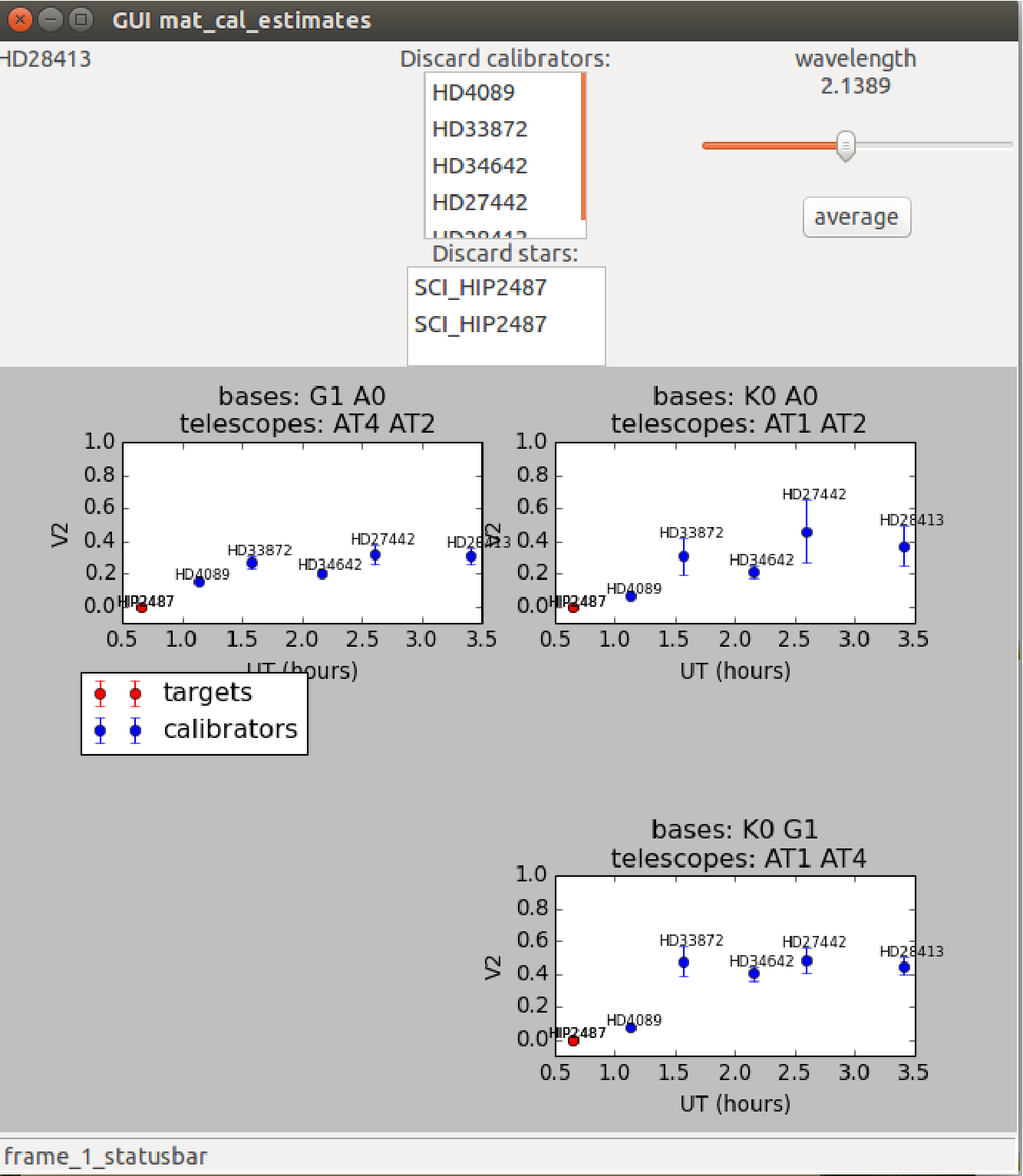}&
\includegraphics[width=0.3\textwidth]{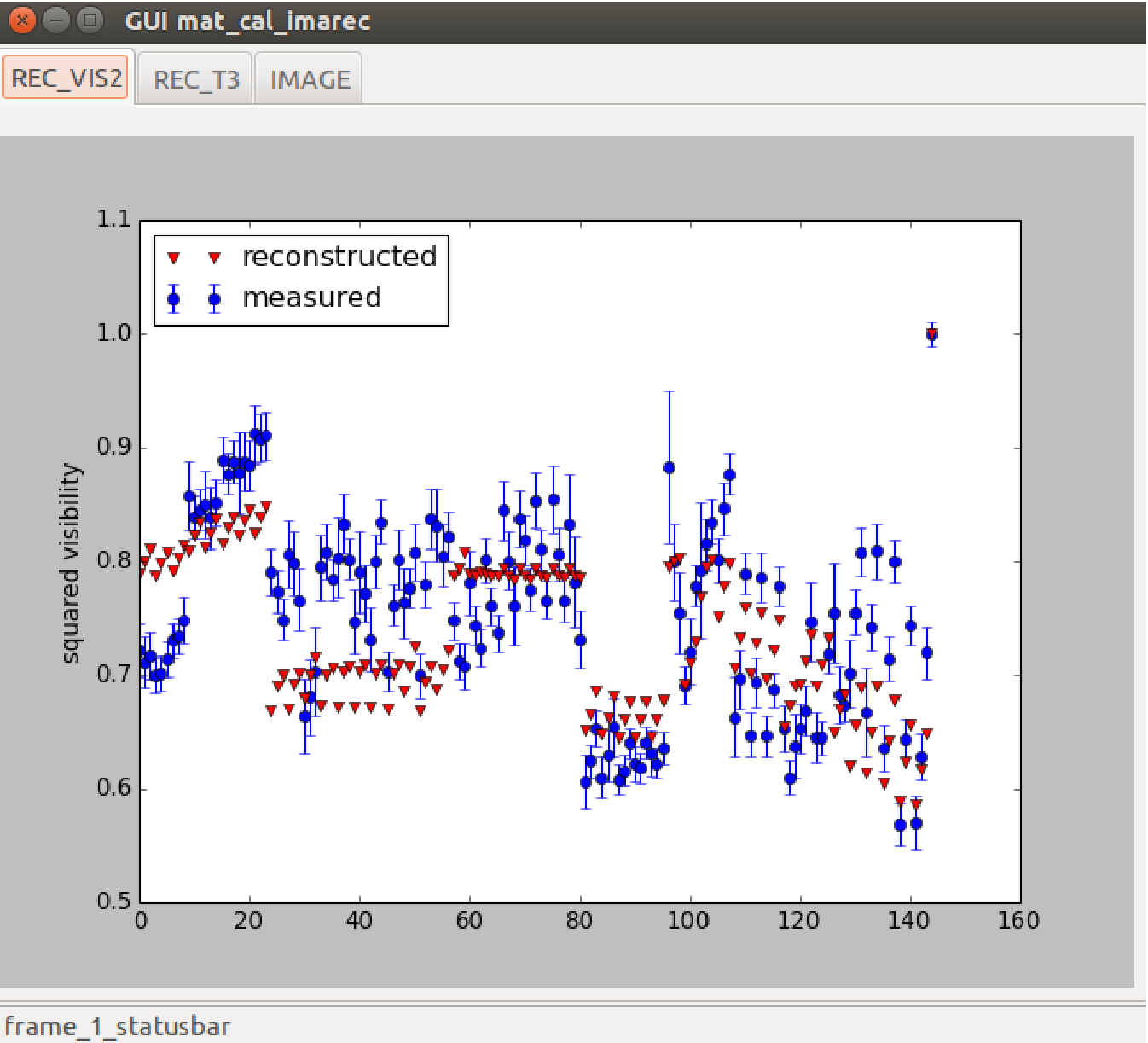}&
\includegraphics[width=0.3\textwidth]{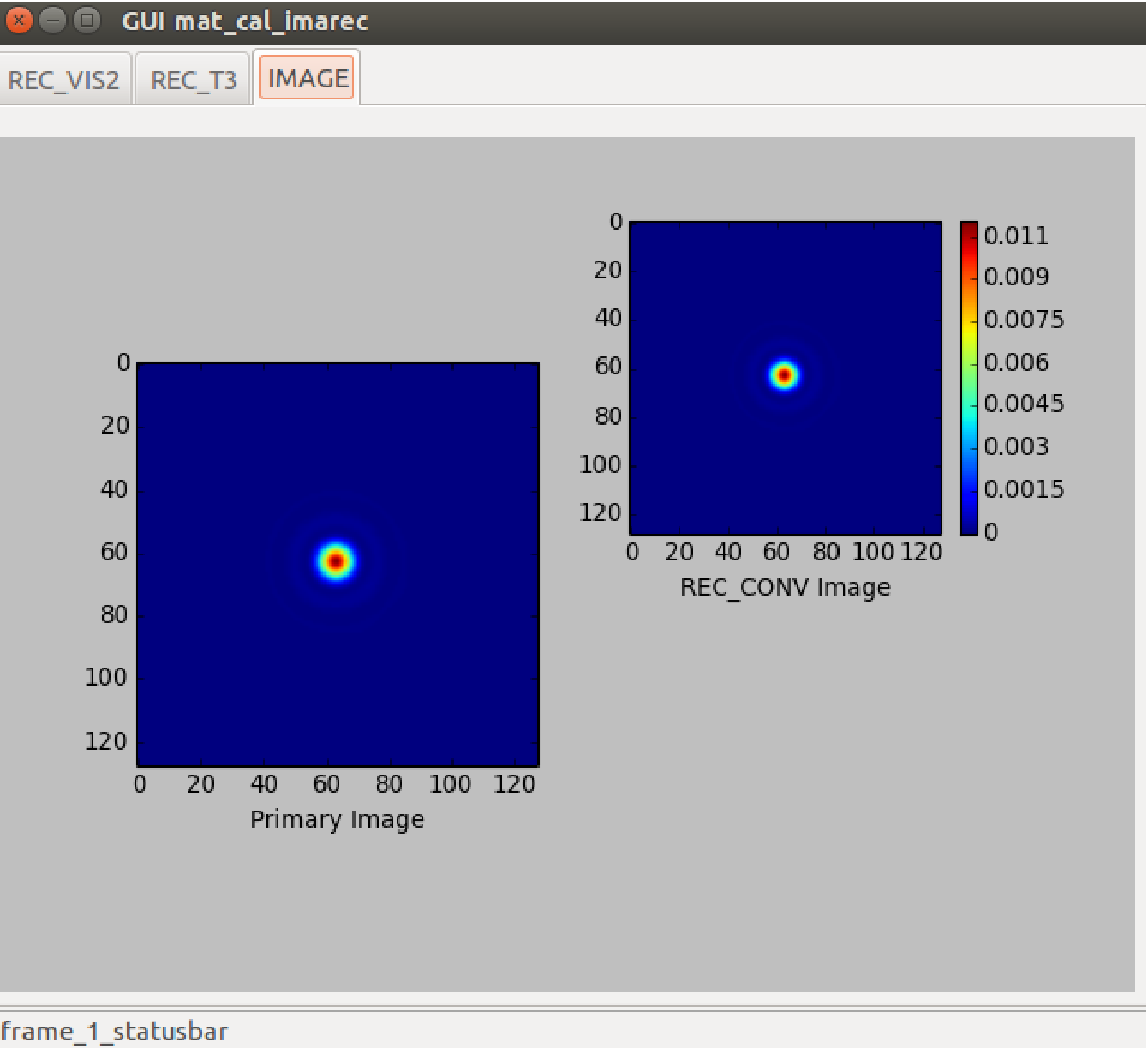}
\end{tabular}
\end{center}
\caption[Kappa Matrix] 
{ \label{fig:pythonactor} Example displays of
the \texttt{drsmat} workflow: {\bf Left:} Transfer function
plot and selection, {\bf Middle:} Visibility comparison plot
between observed (blue) and reconstructed (red)
visibilities, coming from the image reconstruction process,
{\bf Right:} Image display showing the reconstructed image
and the prior image.}
\end{figure}

\texttt{drsmat} uses the ESO-developed \texttt{Reflex} interface to run all the steps, and the
calibration part contains a display of the result, allowing the user
to select the relevant calibration stars (see
Fig.~\ref{fig:pythonactor})

\subsection{Image reconstruction} 

The last step of \texttt{drsmat} is the image reconstruction. MATISSE
will use the IRBIS algorithm, which was specially developed for
it. The content of IRBIS is extensively described in another paper of
this conference\cite{Hofmann2016}. The \texttt{drsmat} software
displays the results of the reconstruction through a python interface,
shown in Fig.~\ref{fig:pythonactor}.

\section{CONCLUSION}

We presented the MATISSE data reduction software \texttt{drsmat} as it
has been implemented for the MATISSE instrument. \texttt{drsmat}
produces calibrated dispersed visibilities, closure phases and
differential phases from the beginning, and also includes the IRBIS
algorithm for image reconstruction. MATISSE is presently undergoing
laboratory tests to verify that it meets all the necessary
requirements to reach its science goals. The tests of \texttt{drsmat}
is a significant part of these tests.

\acknowledgments     

We are grateful to ESO, CNRS/INSU, and the Max-Planck Society for
continuous support in the MATISSE project.


\bibliography{main}   
\bibliographystyle{spiebib}   

\end{document}